\documentclass[a4paper,11pt]{article}

\usepackage{jheppub}

\usepackage[nolist,nohyperlinks]{acronym}
\usepackage{amsmath}
\usepackage{amssymb}
\usepackage{siunitx}
\usepackage{bbm}
\usepackage{booktabs}
\usepackage{braket}
\usepackage{diagbox}
\usepackage{graphicx}
\usepackage{hyperref}
\usepackage[nameinlink,capitalize]{cleveref} 
\usepackage[utf8]{inputenc}
\usepackage{multirow}
\usepackage{placeins}
\usepackage{slashed}
\usepackage[dvipsnames]{xcolor}
\usepackage[normalem]{ulem}
\usepackage{xspace}
\usepackage{subfig}

\allowdisplaybreaks

\DeclareMathOperator{\diag}{diag}

\let\Im\relax
\DeclareMathOperator{\Im}{Im}

\newcommand{\ie}{\textit{i.e.}\xspace}
\newcommand{\eg}{\textit{e.g.}\xspace}

\newcommand{\eps}{\varepsilon}

\renewcommand{\P}{\mathcal{P}}
\newcommand{\GeV}{\text{GeV}}
\newcommand{\SUF}{\ensuremath{\text{SU(3)}_F}\xspace}
\newcommand{\haS}{{\hat{\alpha}_s}}

\newcommand{\EOS}{\texttt{EOS}\xspace}
\newcommand{\vecx}{\vec{x}\xspace}
\newcommand{\FNALMILC}[1][]{\texttt{FNAL/MILC#1}\xspace}
\newcommand{\HPQCD}[1][]{\texttt{HPQCD#1}\xspace}
\newcommand{\JLQCD}[1][]{\texttt{JLQCD#1}\xspace}
\newcommand{\LQCD}{\texttt{LQCD}\xspace}
\newcommand{\QCDSR}{\texttt{QCDSR}\xspace}

\newcommand{\BToD}{\ensuremath{\bar{B}\to D}\xspace}
\newcommand{\BToDstar}{\ensuremath{\bar{B}\to D^*}\xspace}
\newcommand{\BToDDstar}{\ensuremath{\bar{B}\to D^{(*)}}\xspace}
\newcommand{\BstarToD}{\ensuremath{\bar{B}^*\to D}\xspace}
\newcommand{\BstarToDstar}{\ensuremath{\bar{B}^*\to D^*}\xspace}
\newcommand{\BstarToDDstar}{\ensuremath{\bar{B}^*\to D^{(*)}}\xspace}
\newcommand{\BBstarToDDstar}{\ensuremath{\bar{B}^{(*)}\to D^{(*)}}\xspace}

\newcommand{\BsToDs}{\ensuremath{\bar{B}_s\to D_s}\xspace}
\newcommand{\BsToDsstar}{\ensuremath{\bar{B}_s\to D_s^*}\xspace}

\newcommand{\BsstarToDsstar}{\ensuremath{\bar{B}_s^*\to D_s^*}\xspace}

\newcommand{\BqToDq}{\ensuremath{\bar{B}_q\to D_q}\xspace}
\newcommand{\BqToDqStar}{\ensuremath{\bar{B}_q\to D_q^*}\xspace}
\newcommand{\BqToDqDqStar}{\ensuremath{\bar{B}_q\to D_q^{(*)}}\xspace}

\newcommand{\schii}[2]{\chi^{#1}_{#2}}
\newcommand{\schi}[3]{\chi^{#1}_{#2}\big|_{#3}}

\newcommand{\T}[1]{\mathcal{T}\left\lbrace #1 \right\rbrace}

\newcommand{\mB}{m_{B}}
\newcommand{\mD}{m_{D}}
\newcommand{\mBs}{m_{B^*}}
\newcommand{\mDs}{m_{D^*}}

\newcommand{\lBDs}{\lambda_{BD^*}}
\newcommand{\lBsD}{\lambda_{B^*D}}
\newcommand{\lBsDs}{\lambda_{B^*D^*}}

\definecolor{darkgreen}{rgb}{0.0,0.6,0.0}

\newcounter{TODO}

\makeatletter

    \def\dvd{\@ifstar\@@dvd\@dvd}
    \newcommand{\@dvd}[1]{\textcolor{purple}{[\textbf{DvD:} #1]}}
    \newcommand{\@@dvd}[1]{\textcolor{purple}{#1}}
\makeatother

\title{Challenging $\boldsymbol{\bar{B}_{(s)}\to D_{(s)}^{(*)}}$ Form Factors with the Heavy Quark Expansion}

\author[a,b,c]{Marzia Bordone}
\emailAdd{marzia.bordone@cern.ch}
\affiliation[a]{Center for Particle Physics Siegen, Theoretische Physik 1, Universit\"at Siegen
57068 Siegen, Germany}
\affiliation[b]{Theoretical Physics Department, CERN, Geneva, Switzerland}
\affiliation[c]{Physik-Institut, Universit\"at Z\"urich, CH-8057 Z\"urich, Switzerland}

\author[d]{Nico Gubernari}
\emailAdd{nico.gubernari@gmail.com}
\affiliation[d]{DAMTP, University of Cambridge, Cambridge CB3~0WA, UK}

\author[e]{Martin Jung}
\emailAdd{martin.jung@unito.it}
\affiliation[e]{Dipartimento di Fisica, Università di Torino \& INFN, Sezione di Torino, I-10125~Torino, Italy}

\author[f]{Danny van Dyk}
\emailAdd{danny.van.dyk@gmail.com}
\affiliation[f]{Institute for Particle Physics Phenomenology and Department of Physics,\\
Durham University, Durham~DH1~3LE, UK}

\preprint{%
\begin{minipage}{.5\textwidth}
    \raggedleft
    CERN-TH-2025-092, EOS-2025-03, IPPP/25/25,
    P3H-25-032, SI-HEP-2025-10, ZU-TH 35/25
\end{minipage}
}

\abstract{
    Recent publications by three lattice QCD collaborations have provided an unprecedented
    wealth of theoretical predictions for the $\bar{B}_q \to D_q^{(*)}$ form factors, for spectator flavours $q=u/d$ and $q=s$.
    We analyse these predictions within the framework of the heavy-quark expansion (HQE) to order $\alpha_s$, $1/m_b$, and $1/m_c^2$.
    For the first time, our analysis imposes unitarity bounds for all of the $\bar{B}_q^{(*)} \to D_q^{(*)}$ form factors;
    this includes newly identified tensor form factors arising in $\bar{B}_q^*\to D_q^{(*)}$.
    This enables us to treat all form factors in the same fashion.
    At the level of our present analysis, the inclusion of the tensor bounds is not yet constraining the HQE parameter space.
    We find the lattice QCD results to be well compatible with each other in a joint HQE fit as well as
    with QCD sum rule estimates that were used in previous HQE analyses.
    This is in contrast to the strong variability of the posterior predictions, in particular of the form factors ratios $R_0$ and $R_2$.
    Using the posterior distributions of our HQE analysis, we provide predictions for angular observables
    and LFU ratios in the $\bar{B}_q \to D_q^{(*)}\ell^-\bar{\nu}$ decays.
}

\begin{document}

\maketitle

\acrodef{OPE}{Operator Product Expansion}
\newcommand{\OPE}{\ac{OPE}\xspace}
\acrodef{HQE}{Heavy Quark Expansion}
\newcommand{\HQE}{\ac{HQE}\xspace}
\acrodef{SM}{Standard Model}
\newcommand{\SM}{\ac{SM}\xspace}
\acrodef{BSM}{Beyond the Standard Model}
\newcommand{\BSM}{\ac{BSM}\xspace}

\section{Introduction}

The \SM of Particle Physics relies on a small number of free parameters, which are not fixed by
the guiding principles of Poincaré symmetry and gauge invariance. Chief among them are the
Yukawa couplings of the up- and down-type quark fields and their misalignment, generally expressed in terms
of the Cabibbo-Kobayashi-Maskawa (CKM) quark mixing matrix. Consequently, these parameters must be inferred
from experimental data~\cite{PDG:2024cfk,HeavyFlavorAveragingGroupHFLAV:2024ctg}.
This poses a challenge, since QCD confinement prohibits us from accessing the CKM matrix elements
directly from flavour-changing quark currents. In order to extract them, we must rely instead on our understanding of hadronic matrix
elements of said currents, most commonly in the context of low-energy semileptonic processes~\cite{BaBar:2014omp}.
Amongst the CKM matrix elements, the element $V_{cb}$ stands out. Not only in terms of phenomenological importance, given for instance its crucial role in theoretical predictions
for Kaon and rare $b$-hadron decays~\cite{Charles:2004jd,UTfit:2005ras,Buras:2021nns}, but also in terms of experimental precision, since it is amenable to determination from large data sets of semileptonic $b$-hadron decays.
\\

Over the last decade, we have experienced a paradigm shift in the quantity and quality of the theoretical
predictions for the hadronic matrix elements relevant to exclusive $V_{cb}$ determinations. Chief among them are results obtained from
lattice QCD simulations for the hadronic matrix elements governing semileptonic \BqToDqDqStar  transitions~\cite{Bailey:2012rr,MILC:2015uhg,Na:2015kha,McLean:2019qcx,FermilabLattice:2021cdg,Harrison:2023dzh,Aoki:2023qpa}.
Lattice QCD simulations have the potential to provide first-principle results for these hadronic matrix elements,
with quantifiable systematic uncertainties that can to be further improved in future analyses.
In the long run, they are expected to replace the reliance on QCD sum rule calculations~\cite{Neubert:1992wq,Neubert:1992pn,Ligeti:1993hw,Gubernari:2018wyi,Bordone:2019guc}, which have hard-to-quantify systematic uncertainties.
This progress is essential to prepare for the expected high-precision measurements
of exclusive $b\to c\ell\bar\nu$ processes by the LHCb and Belle II experiments.
While the exclusive determinations, particularly from $\bar{B} \to D^* \ell \nu$ decays, have become increasingly precise thanks in large part to lattice QCD inputs, they remain in mild tension (at the $\simeq 2\text{--}3\sigma$ level) with inclusive extractions based on the \OPE~\cite{HeavyFlavorAveragingGroupHFLAV:2024ctg,Fael:2018vsp,Bordone:2021oof,Bernlochner:2022ucr,Finauri:2023kte}.
This problem is further exacerbated by the fact that some lattice QCD predictions 
for the univariate differential distributions in $\bar{B} \to D^* \ell \nu$ decays 
do not agree well with their respective experimental determinations \cite{FermilabLattice:2021cdg,Harrison:2023dzh}.
Understanding and resolving such discrepancies --- chiefly the so-called ``inclusive vs exclusive $|V_{cb}|$ puzzle'' --- is crucial, as it may hint at limitations in the theoretical frameworks or possibly point to physics beyond the Standard Model (BSM).
\\

In two previous analyses~\cite{Bordone:2019vic,Bordone:2019guc}, we have jointly studied the few then-available
lattice QCD results together with QCD sum rule results for \BqToDqDqStar form factors within the framework of the \HQE.
In these studies, the QCD sum rule results were indispensable
due to the large number of fit parameters.
With the availability of lattice QCD results at multiple phase space points and for the full set of \BqToDqStar form factors \cite{Harrison:2023dzh}
this situation has changed. As we show in this work, we can now perform an \HQE analysis of the form factors
in a virtually identical setup to Refs.~\cite{Bordone:2019vic,Bordone:2019guc} while only using lattice QCD inputs.
In this work, we investigate the recent results for the \BqToDqDqStar form factors with the following questions in mind:
\begin{itemize}
    \item Can the wealth of lattice QCD results for these form factors be simultaneously described within the \HQE?
    If not, which of the inputs are in conflict either with each other or with the swift convergence of the \HQE?

    \item Are the lattice QCD results compatible with our previous \HQE results, which were heavily relying on QCD sum rule inputs?

    \item Do the lattice QCD results respect the existing unitarity bounds? Are there further means to constrain
    the \HQE parameters in a model-independent way, beyond what has previously been done in the literature?
\end{itemize}
These questions are designed to test lattice QCD results that are agnostic of the \HQE. As such, they
and our approach to answer them have lasting value beyond the currently available lattice QCD results in the literature.
However, the currently available results are all dependent on some application of an (partially ad-hoc) \HQE to some degree,
typically in the chiral and/or continuum extrapolation stages of the respective analyses~\cite{FermilabLattice:2021cdg,Harrison:2023dzh,Aoki:2023qpa}.
We foresee two possible consequences: (a) problems, \eg regarding the convergence of the \HQE, might be hidden in the final lattice
QCD results and therefore not show up in our study; (b) prior choices for the \HQE parameters used in the
extrapolation stages might show up as spurious problems in our study.
With increasing size of the lattice data sets, we expect that the impact of and necessity for using the \HQE on the lattice will diminish.
\\

Our analysis is complementary to the ones published in Ref.~\cite{Martinelli:2023fwm,Bordone:2024weh} on two accounts, compare also
an earlier analysis of a subset of the lattice QCD results in Ref.~\cite{Ray:2023xjn}.
First, we work strictly within the \HQE. When considering the full set of \BqToDqDqStar form factors,
the \HQE parametrisation is more predictive than the BGL parametrisation used in Ref.~\cite{Bordone:2024weh} and the dispersive matrix method employed in Ref.~\cite{Martinelli:2023fwm}.
Hence, the question as to the mutual compatibility of the
lattice QCD results might be answered differently in our framework.
Second, in Ref.~\cite{Bordone:2024weh}, form factor parameters are inferred either from theory only or theory and experimental measurements of the $\BToDstar \ell^-\bar\nu$ angular distribution.
Here, we use theory information only, since using experimental measurements
would complicate the analysis further: by introducing the CKM matrix element $V_{cb}$, and on account of
the inability to separate $V_{cb}$ from the normalisation of the form factors without theory input on the form factors.
We leave these analyses for future work.

\section{Theoretical framework}
\label{sec:theory}

This work focuses on $\bar{B}_q(p)\to D_q^{(*)}(k)$ transitions, which can be described in terms of ten independent form factors per spectator flavour $q$ that depend on the momentum transfer squared $q^2 = (p-k)^2$.%
\footnote{%
For simplicity, we suppress the spectator flavour index in this section, which applies universally to all meson states and properties, without adopting a flavour-symmetry assumption at this stage. In fact, while we do assume isospin symmetry throughout the phenomenological analysis, SU(3)-flavour (\SUF) symmetry is not assumed unless explicitly stated.
}
For the description of $\BToDDstar \ell^-\bar\nu$ processes in the \SM, it suffices to discuss the three form factors
arising from the vector current ($f_+$, $f_0$, $V$) and the three form factors arising from the axial current ($A_1$, $A_{12}$, $A_0$),
with $f_0$ and $A_0$ doubling as the form factors of the scalar and pseudoscalar currents, respectively.
Throughout this work, we use and display our numerical results for the basis detailed in \cref{sec:FF-def} and Refs.~\cite{Bordone:2019guc,Bordone:2019vic}.\\

In this section, we focus instead on the form factors arising from the tensor currents.
Although these form factors are only \emph{needed} in a BSM setting with tensor operators,
they nevertheless provide important independent information on the heavy-quark-expansion parameters because of heavy-quark spin symmetry.
This is possible, since the lattice results for the tensor form factor are largely independent
of the results for the vector and axial form factors.
As a consequence, including the tensor form factors, even in an \SM analysis of the data, will reduce the overall theoretical uncertainties.
To this end, we discuss both a \HQE and a BGL-like basis of \BToDDstar tensor form factors, including their saturations
of the unitarity bounds.
In addition, we define for the first time a basis of the \BstarToDDstar tensor form factors.
We derive their relation to the \HQE parameters of \BToDDstar form factors and calculate their contributions to the saturation of the unitary bounds.
The latter requires a BGL-inspired parametrisation of the \BstarToDDstar tensor form factors, which we also present for the first time.
Our results ensure that the saturations of the unitarity bounds for all \BBstarToDDstar form factors can be consistently used in phenomenological analyses.

\subsection{Form factor definitions} 
\label{sec:FF-def}

In full generality, we can describe the hadronic matrix elements of  $\bar B^{(*)}(p)\to D^{(*)}(k)$ transitions mediated by the tensor current or axial tensor current in terms of 14 form factors.
Employing the same conventions as in Refs.~\cite{Bordone:2019guc,Bordone:2019vic}, we have for the $\bar B(p)\to D^{(*)}(k)$ transitions
\begin{align}
    \langle D(k)|\bar{c}\sigma^{\mu\nu}q_\nu b|\bar{B}(p)\rangle =&\, \frac{i}{m_B+m_D}\left[2 q^2p^\mu-(m_B^2-m_D^2+q^2)q^\mu\right]f_T  \,, \label{eq:BtoD-trad1}\\
    \langle D^*(k,\eta)|\bar{c}\sigma^{\mu\nu}q_\nu b|\bar{B}(p)\rangle =&\,-2 i \epsilon^{\mu\nu\alpha\beta} \eta_\nu^{*}p_\alpha k_\beta T_1 \,, \label{eq:BtoDst-trad1}\\
    \langle D^*(k,\eta)|\bar{c}\sigma^{\mu\nu}q_\nu \gamma_5 b|\bar{B}(p)\rangle =&\, \eta_\alpha^{*} \bigg\{\left[(m_B^2-\mDs^2)g^{\mu\alpha}-p^\alpha (p+k)^\mu\right]T_2 
    \nonumber\\*
    &+p^\alpha\left[q^\mu-\frac{q^2}{m_B^2-\mDs^2}(p+k)^\mu\right]T_3\bigg\} \,. \label{eq:BtoDst-trad2}
\end{align}
Here, $\eta$ is the polarization vector of the $D^*$ meson.
Note that throughout this work we keep the $q^2$ dependence of the form factors implicit unless explicitly required.
The helicity form factors --- which diagonalize the unitarity bounds --- are $T_1$, $T_2$ and $T_{23}$. The latter is defined as
\begin{equation}
    T_{23} = \frac{(m_B^2-\mDs^2)(m_B^2+3 \mDs^2-q^2)T_2-\lBDs T_3}{8 m_B \mDs^2(m_B-\mDs)}\,, \label{eq:BtoDst-trad3}
\end{equation}
where we abbreviate the frequently occurring Källén function 
\begin{equation}
    \lambda_{B^{(*)}D^{(*)}} \equiv \lambda(m_{B^{(*)}}^2,m_{D^{(*)}}^2,q^2)=(m_{B^{(*)}}^2 - m_{D^{(*)}}^2 -q^2)^2-4m_{D^{(*)}}^2q^2\,.
\end{equation}
For the $\bar B^{*}(p)\to D(k)$ transition, we find the following to be a useful decomposition of the hadronic matrix elements in terms of form factors:
\begin{align}
    \langle D(k)|\bar{c}\sigma^{\mu\nu} q_\nu b|\bar{B}^*(p,\varepsilon)\rangle &=
        \,  2 i\eps^{\mu\nu\alpha\beta} \varepsilon_\nu k_\alpha p_\beta \bar{T}_1  \,,  \label{eq:BsttoD-trad1}
        \\
        \langle D(k)|\bar{c}\sigma^{\mu\nu}q_\nu \gamma_5 b|\bar{B}^*(p,\varepsilon)\rangle &=\, \varepsilon_\alpha \bigg\{\left[(\mBs^2-\mD^2)g^{\mu\alpha}+k^\alpha (p+k)^\mu\right]\bar{T}_2 
        \nonumber\\*
        &\phantom{=}+k^\alpha\left[q^\mu -\frac{q^2}{\mBs^2-\mD^2}(p+k)^\mu\right]\bar{T}_3\bigg\} \,.\label{eq:BsttoD-trad2}
\end{align}
Here, $\varepsilon$ is the polarization vector of the $\bar{B}^*$ meson.
Notice that the definitions in \cref{eq:BsttoD-trad1,eq:BsttoD-trad2} resemble closely the ones for $\bar B(p)\to D^{*}(k)$ tensor form factors.
This comes about, because the $\bar B(p)\to D^{*}(k)$ and $\bar B^*(p)\to D(k)$ form factors are connected by the combination
of crossing symmetry and exchange of the momenta and masses of the heavy quarks and hadrons.
Similar to the $\bar B(p)\to D^{*}(k)$ case, we need to introduce the redefinition
\begin{equation}
    \bar{T}_{23} = \frac{(\mBs^2-\mD^2)(3\mBs^2+\mD^2-q^2)\bar{T}_2+\lBsD \bar{T}_3}{8 \mBs^2 \mD (\mBs-\mD)}\label{eq:BsttoD-trad3}
\end{equation}
to obtain the helicity base.
For the $\bar{B}^{*}(p)\to D^*(k)$ transition, we find the following to be a useful decomposition of the hadronic matrix elements in terms of form factors:
\begin{align}
    \langle D^*(k,\eta)|&\bar{c}\sigma^{\mu\nu}q_\nu\gamma_5b|\bar{B}^*(p,\epsilon)\rangle = 
    \nonumber\\*
        &\,+\frac{1}{\lBsDs}\epsilon_\alpha \eta^*_\beta\bigg\{2\mBs\mDs^2 p^\beta k_\nu p_\lambda \eps^{\mu\nu\lambda\alpha}T_5+ 2\mBs^2\mDs k^\alpha k_\nu p_\lambda \eps^{\mu\nu\lambda\beta}T_6\notag \\*
        &
        -\frac{s_+}{\sqrt{\mBs\mDs}}
        [(\mBs^2-\mDs^2+q^2)k^\alpha k_\nu p_\lambda \eps^{\mu\nu\lambda\beta}+(\mBs^2-\mDs^2-q^2)p^\beta k_\nu p_\lambda \eps^{\mu\nu\lambda\alpha}\notag\\*
        &-\frac{\lBsDs}{2}q_\nu \eps^{\mu\nu\alpha\beta}]T_4\bigg\}\,,  \label{eq:BstToDst-trad1}\\
     \langle D^*(k,\eta)|&\bar{c}\sigma^{\mu\nu}q_\nu b|\bar{B}^*(p,\epsilon)\rangle=
    \nonumber\\
        &\,\frac{i\epsilon_\alpha \eta^*_\beta}{\lBsDs}
        \bigg\{\mBs k^\alpha\left[\lBsDs g^{\beta\mu}-2p^\beta(k^\mu(\mBs^2+\mDs^2-q^2)-2\mDs^2 p^\mu)\right]T_8\notag\\
        &+ \mDs p^\beta\left[2 k^\alpha (2\mBs^2 k^\mu-(\mBs^2+\mDs^2-q^2)p^\mu)+\lBsDs g^{\alpha\mu}\right]T_9
        \notag\\*
        &-2\sqrt{\mBs\mDs}k^\alpha p^\beta\left[(\mBs^2-\mDs^2+q^2)k^\mu-(\mBs^2-\mDs^2-q^2)p^\mu\right]T_{10}
        \notag\\*
        &-\frac{1}{2 \sqrt{\mBs \mDs}}
        [
            (\lBsDs g^{\alpha\beta} -2k^\alpha p^\beta(\mBs^2+\mDs^2-q^2))
        \notag\\*
       & \times
            (k^\mu(\mBs^2-\mDs^2+q^2)- p^\mu(\mBs^2-\mDs^2-q^2))
       ]T_7\bigg\}\,, 
       \label{eq:BstToDst-trad2}
\end{align}
where the form factors $T_4$ through $T_{10}$ are the helicity form factors,
and we abbreviate $s_\pm = t_\pm-q^2$ with $t_\pm \equiv (m_{B^{(*)}} \pm m_{D^{(*)}})^2$.
\\

As in the case of the \BToDDstar form factors, the newly defined \BstarToDDstar form factors also feature so-called endpoint relations, \ie, exact relations
between the form factors that arise from algebraic identities and equations of motion.
We derive the following endpoint relations:
\begin{align}
    \bar{T}_1(q^2=0)&=\bar{T}_2(q^2=0)\,, \\
    T_9(q^2=0) &= \frac{\mBs\mDs}{\mBs^2-\mDs^2}T_5(q^2=0)\,,\\T_8(q^2=0) &= \frac{\mBs\mDs}{\mBs^2-\mDs^2}T_6(q^2=0)\,,\\
    T_5(q^2=t_-) &= T_6(q^2=t_-) =4 \frac{\mBs-\mDs}{\sqrt{\mBs\mDs}}T_4(q^2=t_-)\,,\\ T_{10}(q^2=t_-) &= T_7(q^2=t_-) +\frac{\sqrt{\mBs\mDs}}{\mBs-\mDs}\left(T_9(q^2=t_-)-T_8(q^2=t_-)\right)\,.
\end{align}
The first equation is analogous to the identity $T_1(0) = T_2(0)$.
All remaining relations are required to avoid spurious kinematic singularities in the definitions \cref{eq:BstToDst-trad1,eq:BstToDst-trad2} due to zeros of the K\"all\'en function in the denominator. Note that all form factors defined here are dimensionless.

\subsection{Heavy-quark expansion} 
\label{sec:theory:hqe}

Within the \HQE, each \BBstarToDDstar form factor $F$ is expanded in a triple series,
with expansion parameters $\varepsilon_Q \equiv \bar\Lambda/2 m_Q$, where $Q=c,b$, and $\haS \equiv \alpha_s/\pi$.\footnote{
    The parameter $\bar\Lambda$ is appearing also in the heavy meson mass expansions in HQE, and is of the order $\bar\Lambda\sim 500\,\mathrm{MeV}$.
    It is specific to the spectator quark flavour $q$.
}
Following Refs.~\cite{Bordone:2019guc,Bordone:2019vic}, we adopt a power counting $\varepsilon_b \sim \varepsilon_c^2 \sim \haS \sim \varepsilon^2$
and work to order $\varepsilon^2$.
The terms in this expansion are expressed as products of calculable coefficient functions of the kinematics and $\alpha_s$,
and non-perturbative matrix elements, the so-called Isgur-Wise functions~\cite{Isgur:1989vq,Isgur:1990yhj}.
For the vector and axial form factors, this is discussed in great detail in the literature~\cite{Falk:1992wt,Caprini:1997mu,Bernlochner:2017jka}.
For these form factors we use the conventions of Ref.~\cite{Bordone:2019vic}.
For the tensor form factors of \BToDDstar transitions, this expansion was discussed for the first time up to order $\varepsilon_Q$ in Ref.~\cite{Bernlochner:2017jka} and to partial $\varepsilon_Q^2$ in Ref.~\cite{Bernlochner:2022ywh}.
\\

For the purpose of the \HQE, it is convenient to use the following basis of the hadronic form factors\footnote{
    We adopt the relativistic normalization of states, meaning our states have a mass dimension of $-1$.
}
for the \BToDDstar form factors:
\begin{align}
        \langle D(k)|&\bar{c}\sigma^{\mu\nu} q_\nu b|\bar{B}(p)\rangle = 
        \frac{i}{2\sqrt{m_Bm_D}}\left[2 q^2p^\mu-(m_B^2-m_D^2+q^2)q^\mu\right] h_T
        \,, \label{eq:BtoDtensor1} \\
        \langle D^*(k,\eta)|&\bar{c}\sigma^{\mu\nu} q_\nu b|\bar{B}(p)\rangle = 
        \, \frac{i}{ \sqrt{m_Bm_{D^*}}} \eps^{\mu\nu\alpha\beta} k_\alpha p_\beta \eta_\nu^* [(\mB+\mDs)h_{T_1}-(\mB-\mDs)h_{T_2}]
        \,. \label{eq:BtoDstensor1}\\
        \langle D^*(k,\eta)|&\bar{c}\sigma^{\mu\nu} q_\nu \gamma_5 b|\bar{B}(p)\rangle = 
        \nonumber\\*
        &\, 
        -\frac{1}{2\sqrt{m_{D^*}m_B^3}}\eta_\nu^* 
        \bigg\{m_B g^{\mu\nu}[(\mB-\mDs)s_+ h_{T_1} -(\mB+\mDs) s_-h_{T_2}]\notag\\
        & +k^\mu p^\nu[2m_B^2 (h_{T_2}-h_{T_1})-(\mB^2-\mDs^2+q^2) h_{T_3} ]  
        \notag \\
        & - p^\mu p^\nu[2 \mDs \mB (h_{T_1}+h_{T_2})-(\mB^2-\mDs^2-q^2)h_{T_3} ] \bigg\} \,, \label{eq:BtoDstensor2}
\end{align}
For the \BstarToD form factors, we introduce
\begin{align}
        \langle D(k)|&\bar{c}\sigma^{\mu\nu} q_\nu b|\bar{B}^*(p,\varepsilon)\rangle =
        \, \frac{-i}{ \sqrt{m_{B^*}m_D}} \eps^{\mu\nu\alpha\beta} p_\alpha k_\beta \varepsilon_\nu [(\mBs+m_D)h_{\bar{T}_1}+(\mBs-m_{D})h_{\bar{T}_2}]\,, \label{eq:BstoDtensor1} \\
         \langle D(k)|&\bar{c}\sigma^{\mu\nu}q_\nu\gamma_5b|\bar{B}^*(p,\varepsilon)\rangle =
    \nonumber\\
        &\, -\frac{1}{2\sqrt{m_{B^*}m_D^3}}\varepsilon_\nu \bigg\{m_D g^{\mu\nu}[(\mBs-m_D)s_+ h_{\bar{T}_1} +(\mBs+m_D) s_-h_{\bar{T}_2}]\notag\\
        & +p^\mu k^\nu[2m_D^2 (h_{\bar{T}_1}-h_{\bar{T}_2})-(\mBs^2-m_D^2-q^2) h_{\bar{T}_3} ]  \notag \\
        & + k^\mu k^\nu[2 m_D \mBs (h_{\bar{T}_1}+h_{\bar{T}_2})+(\mBs^2-m_D^2+q^2)h_{\bar{T}_3} ] \bigg\} \,. \label{eq:BstoDtensor2}
\end{align}
For the \BstarToDstar form factors, we introduce
\begin{align}
    \langle D^*(k,\eta)|&\bar{c}\sigma^{\mu\nu}q_\nu b|\bar{B}^*(p,\varepsilon)\rangle=
    \nonumber\\*
        &\,\frac{i}{2\sqrt{\mBs\mDs}}\varepsilon_\alpha\eta^*_\beta\left\{g^{\alpha\beta} h_{T_7}\left[(\mBs^2-\mDs^2-q^2)p^\mu-(\mBs^2-\mDs^2+q^2)k^\mu\right]\right. \notag \\
        &+k^\alpha g^{\beta\mu}\left[h_{T_6}(\mBs^2-\mDs^2+q^2)-\frac{\mBs}{\mDs}h_{T_8}(\mBs^2-\mDs^2-q^2)+ s_+ h_{T_4}\right] \notag\\
        &+p^\beta g^{\alpha\mu}\left[-h_{T_5}(\mBs^2-\mDs^2-q^2)-\frac{\mDs}{\mBs}h_{T_9}(\mBs^2-\mDs^2+q^2)+s_+ h_{T_4}\right]\notag\\
        &+2 p^\beta  k^\alpha\left[k^\mu \left(\frac{\mBs}{\mDs}h_{T_8}-h_{T_5}-\frac{\mBs^2-\mDs^2+q^2}{2\mBs\mDs}h_{T_{10}}\right)\right. \notag \\
        &\left.\left.-p^\mu\left(h_{T_6}+\frac{\mDs}{\mBs}h_{T_9}-\frac{\mBs^2-\mDs^2-q^2}{2\mBs\mDs}h_{T_{10}}\right)\right]\right\}\,,\label{eq:BstoDstensor1} \\
    \langle D^*(k,\eta)|&\bar{c}\sigma^{\mu\nu}q_\nu\gamma_5b|\bar{B}^*(p,\varepsilon)\rangle = 
    \nonumber\\*
        &\, \frac{-1}{2\sqrt{\mBs\mDs}}\varepsilon_\alpha\eta^*_\beta\left[-2\left(h_{T_6}-\frac{\mBs}{\mDs}h_{T_8}\right)\eps^{\mu\nu\lambda\beta} k_\nu p_\lambda k^\alpha \right. 
        \notag \\*
        &\left.
        +2\left(\frac{\mDs}{\mBs}h_{T_9}+h_{T_5}\right)\eps^{\mu\nu\lambda\alpha}k_\nu p_\lambda p^\beta
        - h_{T_4}s_+\eps^{\mu\nu\alpha\beta}q_\nu\right]
        \,,
        \label{eq:BstoDstensor2}
\end{align}
where $s_\pm = t_\pm-q^2$, and  $t_\pm \equiv (m_{B^{(*)}} \pm m_{D^{(*)}})^2$. The appropriate masses for a given form factor in these expressions are those of the corresponding initial and final-state mesons.
The change of basis between \cref{eq:BtoDtensor1,eq:BtoDstensor1,eq:BtoDstensor2,eq:BstoDtensor1,eq:BstoDtensor2,eq:BstoDstensor1,eq:BstoDstensor2} and \cref{eq:BtoD-trad1,eq:BtoDst-trad1,eq:BtoDst-trad2,eq:BtoDst-trad3,eq:BsttoD-trad1,eq:BsttoD-trad2,eq:BsttoD-trad3,eq:BstToDst-trad1,eq:BstToDst-trad2}
is detailed in \cref{app:form-factor-decomposition}.
In the Heavy-Quark limit $m_{b,c}\to \infty$ all form factors are proportional to a single function $\xi$, the leading power (LP) Isgur-Wise function~\cite{Isgur:1989vq,Isgur:1990yhj}:
\begin{equation}
\begin{aligned}
    & h_T = h_{T_1} = h_{\bar T_1}  = h_{T_4} = h_{T_5} = h_{T_6} = h_{T_7} = \xi(w)\,, \\
    & h_{T_2} =h_{T_3} = h_{\bar T_2} = h_{\bar T_3} = h_{T_8} =h_{T_9} = h_{T_{10}} = 0 \,.
\end{aligned}
\end{equation}
It is common to express $\xi$ as a function of
\begin{align}
    \label{eq:wdef}
    w = v\cdot v' = \frac{m_{B^{(*)}}^2  + m_{D^{(*)}}^2 - q^2}{2m_{B^{(*)}} m_{D^{(*)}}}\,.
\end{align}
Throughout this paper we keep the $w$ dependence of the Isgur-Wise functions and HQET Wilson coefficients implicit unless explicitly required.
\\

The corrections to these results at next-to-leading order (NLO) in $\alpha_s$ are obtained from the matching of HQET onto QCD~\cite{Falk:1990yz,Falk:1990cz,Neubert:1992qq},
\begin{multline}
    \bar{c}\sigma^{\mu\nu} b\to \bar{c}_{v^\prime}\left[(1+\haS C_{T_1})\sigma^{\mu\nu}+\haS C_{T_2}i(v^\mu\gamma^\nu-v^\nu\gamma^\mu)
    \right.
    \\*
    \left.
    +\haS C_{T_3}i (v^{\prime\mu}\gamma^\nu-v^{\prime\nu}\gamma^\mu)+\haS C_{T_4}i(v^{\prime\mu}v^\nu-v^{\prime\nu}v^\mu)\right]b_v
    +\mathcal{O}(\haS^2)
    \,,
\end{multline}
where $v_\mu = p_\mu/m_{B^{(*)}}$, $v_\mu' = k_\mu/m_{D^{(*)}}$,
the quark fields on the right-hand side are HQET fields labelled by their four-velocities,
the coefficients $C_{T_i}$ are pertubative coefficients from matching HQET onto QCD,
and $C_{T_4}=0$ at this order.
Introducing next-to-leading power (NLP) and next-to-next-to-leading power (NNLP) terms is done following the same approach as in Ref.~\cite{Bernlochner:2017jka}. 
We find it convenient to factor out the leading-power Isgur-Wise function,
following Refs~\cite{Bernlochner:2017jka,Bordone:2019guc,Bordone:2019vic}. To this end, we define
\begin{equation}
    \hat{h}_i = {h}_i / \xi\,;
\end{equation}
we extend this notation also to the Isgur-Wise functions at NLP and NNLP.
Including power corrections up to order $\mathcal{O}(\varepsilon_c^2)$, the $\BToDDstar$ form factors read~\cite{Bernlochner:2017jka}\footnote{
    We use the conventions of Ref.~\cite{Manohar:2000dt} for time reversal and charge conjugation, adapting the expression of Ref.~\cite{Bernlochner:2017jka} for the $\varepsilon_Q$ corrections.
}
\begin{align}
    \label{eq:hTexp}
    \hat h_T=&\,1+\haS\left(
        C_{T_1}-C_{T_2}+C_{T_3}
    \right)
    +(\varepsilon_c + \varepsilon_b)(\hat L_1 - \hat L_4)
    +\varepsilon_c^2(\hat\ell_1 - \hat\ell_4)\,,\\
    \label{eq:hT1exp}
    \hat h_{T_1}=&\,1+\haS\left(C_{T_1}+\frac{w-1}{2}(C_{T_2}-C_{T_3})\right)+\varepsilon_c\hat L_2+\varepsilon_b \hat L_1+\varepsilon_c^2\hat\ell_2\,,\\
    \label{eq:hT2exp}
    \hat h_{T_2}=&\,\haS\frac{w+1}{2}\left(C_{T_2}+C_{T_3}\right)+\varepsilon_c\hat L_5-\varepsilon_b \hat L_4+\varepsilon_c^2\hat\ell_5\,,\\
    \label{eq:hT3exp}
    \hat h_{T_3}=&\,\haS C_{T_2}+\varepsilon_c(\hat L_6-\hat L_3) +\varepsilon_c^2(\hat\ell_6-\hat\ell_3)\,.
\end{align}
For the \BstarToD form factors, we obtain similarly
\begin{align}
    \label{eq:hT1bexp}
    \hat h_{\bar T_1}=&\,1+\haS\left(C_{T_1}+\frac{w-1}{2}(C_{T_2}-C_{T_3})\right)+\varepsilon_b\hat L_2+\varepsilon_c \hat L_1+\varepsilon_c^2\hat\ell_1\,,\\
    \label{eq:hT2bexp}
    \hat h_{\bar T_2}=\,&-\haS\frac{w+1}{2}\left(C_{T_2}+C_{T_3}\right)+\varepsilon_b\hat L_5-\varepsilon_c \hat L_4-\varepsilon_c^2\hat\ell_4\,,\\
    \label{eq:hT3bexp}
    \hat h_{\bar T_3}=&\,-\haS C_{T_3}+\varepsilon_b(\hat L_6-\hat L_3)\,.
\end{align}
They are related to the \BToDstar ones using crossing symmetry and exchange of the heavy-quark momenta and masses.
Finally, for the \BstarToDstar form factors, we obtain
\begin{align}
    \label{eq:hT4exp}
    \hat h_{T_4}=&\, 1+\haS C_{T_1}+ (\varepsilon_b+\varepsilon_c)\left(\hat{L}_2-\frac{w-1}{1+w}\hat{L}_5\right)+\varepsilon_c^2\left(\hat{\ell}_2-\frac{w-1}{1+w}\hat{\ell}_5\right) \,, \\
    \label{eq:hT5exp}
    \hat h_{T_5}=&\, 1+\haS (C_{T_1}+C_{T_3})+\varepsilon_b (\hat{L}_2-\hat{L}_5)+\varepsilon_c (\hat{L}_2+\hat{L}_6-\hat{L}_3-\hat{L}_5)\\
        &+\varepsilon_c^2 (\hat{\ell}_2+\hat{\ell}_6-\hat{\ell}_3-\hat{\ell}_5)\,, \\
    \label{eq:hT6exp}
    \hat h_{T_6}=&\, 1+\haS (C_{T_1}-C_{T_2})+\varepsilon_b (\hat{L}_2+\hat{L}_6-\hat{L}_3-\hat{L}_5)+\varepsilon_c (\hat{L}_2-\hat{L}_5)+\varepsilon_c^2 (\hat{\ell}_2-\hat{\ell}_5)\,,\\
    \label{eq:hT7exp}
    \hat h_{T_7}=&\, 1+\haS (C_{T_1}-C_{T_2}+C_{T_3}+(w+1)C_{T_4})+(\varepsilon_b+\varepsilon_c) (\hat{L}_2-\hat{L}_5)+\varepsilon_c^2(\hat{\ell}_2-\hat{\ell}_5) \,, \\
    \label{eq:hT8exp}
    \hat h_{T_8}=&\, +\haS C_{T_3}-\varepsilon_b (\hat{L}_3+\hat{L}_6) \,, \\
    \label{eq:hT9exp}
    \hat h_{T_9}=&\, +\haS C_{T_2}+\varepsilon_c (\hat{L}_3+\hat{L}_6)+\varepsilon_c^2 (\hat{\ell}_3+\hat{\ell}_6) \,, \\*
    \label{eq:hT10exp}
    \hat h_{T_{10}} =& -\haS C_{T_4}\,.
\end{align}
The six functions emerging at NLP, $\hat{L}_i$, are not independent due to constraints imposed by the equations of motion.
They can be expressed in terms of three independent NLP-Isgur-Wise functions $\hat\chi_2$, $\hat\chi_3$, and $\eta$ through the following relations~\cite{Luke:1990eg,Falk:1992wt,Bernlochner:2017jka}:
\begin{equation}
\begin{aligned}
    \hat{L}_1 &= - 4(w-1) \hat\chi_2 + 12 \hat\chi_3\,,&
    \hat{L}_2 &= - 4 \hat\chi_3\,, &
    \hat{L}_3 &= 4 \hat\chi_2\,, \\*
    \hat{L}_4 &= 2 \hat\eta - 1 \,, &
    \hat{L}_5 &= -1\,, &
    \hat{L}_6 &= - 2 (1 + \hat\eta)/(w+1)\,,
\end{aligned}
\end{equation}
where a fourth NLP-Isgur-Wise function $\chi_1$ has been absorbed into the leading function $\xi$~\cite{Bernlochner:2017jka}.
Six further functions $\hat{\ell}_i$ emerge at order $\varepsilon_c^2 \sim \varepsilon^2$, \ie, at NNLP both in $\varepsilon_Q$ and in our power counting.
However, they do not form the full set of functions emerging at $\varepsilon_Q^2 = \varepsilon_c^2,\varepsilon_b^2 ,\varepsilon_b\varepsilon_c$~\cite{Falk:1992wt};
the remaining functions enter at order $\varepsilon_b \varepsilon_c \sim \varepsilon^3$ or $\varepsilon_b^2 \sim \varepsilon^4$,
\ie, beyond the accuracy that we aim for.
Although equations of motions apply to the full set of functions entering at $\varepsilon_Q^2$ and therefore reduce the number of independent functions,
we can treat the restricted set at order $\varepsilon_c^2$ as fully independent.
To order $\mathcal{O}(\alpha_s,\varepsilon_b,\varepsilon_c^2)$, the Wilson coefficient $C_{T_4}$ vanishes,
hence, $h_{T_{10}} = 0$ at this order.

\subsection{Unitarity bounds}
\label{sec:th-bounds}

We derive unitarity bounds for the tensor form factors in $\bar{B}^{(*)}\to D^{(*)}$ and $\bar{B}_s^{(*)}\to D_s^{(*)}$ processes.
While most of the formulae are explicitly presented for the $\bar{B}^{(*)}\to D^{(*)}$ case, their extension to $\bar{B}_s^{(*)}\to D_s^{(*)}$ is straightforward.
Following Ref.~\cite{Boyd:1997kz}, we define the two-point correlator
\begin{align}
    \label{eq:boundscorr}
    \Pi^{\mu\nu}_{\Gamma}(q)
        & \equiv i\! \int\! d^4x\, e^{iq\cdot x} \Braket{0 | \T {J_\Gamma^{\mu}(x) J_\Gamma^{\dagger,\nu}(0)} | 0}\,.
\end{align}
We consider the following quark currents $J_\Gamma^\mu$:
\begin{equation}
\begin{aligned}
    J_V^\mu & = \bar{c}\, \gamma^\mu b    \,,&\qquad
    J_A^\mu & = \bar{c}\, \gamma^\mu \gamma_5 b      \,,\\
    J_T^\mu & = \bar{c}\, \sigma^{\mu \alpha}q_\alpha  b     \,,&
    J_{AT}^\mu & = \bar{c}\, \sigma^{\mu \alpha}q_\alpha \gamma_5  b     
    \,.
\end{aligned}
\end{equation}
We also define the scalar correlators $\Pi^{J}_{\Gamma}$ as
\begin{align}
    \label{eq:Pilambda}
    \Pi_\Gamma^J(q^2) 
    =
    \P_{\mu\nu}^{J}(q) \,
    \Pi^{\mu\nu}_{\Gamma}(q)
    \,,
\end{align}
where
\begin{align}
    \P_{\mu\nu}^{1}(q) & = \frac{1}{ (d-1)} \left(\frac{q_\mu q_\nu}{q^2} - g_{\mu\nu}\right)\,,&&
    \P_{\mu\nu}^{0}(q)  = \frac{q_\mu q_\nu}{q^2}\,.
\end{align}
and $d$ is the number of spacetime dimensions.
The functions $\Pi^J_{\Gamma}$ satisfy subtracted dispersion relations:
\begin{align}
    \label{eq:chisub}
    \schii{J}{\Gamma}(Q^2)
        & = \frac{1}{u!} \left[\frac{\partial}{\partial q^2}\right]^u \Pi_\Gamma^{J}(q^2)
        \bigg|_{q^2=Q^2}
          = \frac{1}{\pi} \int\limits_0^\infty dt\, \frac{\Im \Pi_\Gamma^{J}(t)}{(t - Q^2)^{u+1}}\,,
    \qquad \text{with }J=0,1. 
\end{align}
Two representations of \cref{eq:chisub} are required for the construction of the unitarity bounds:
a partonic representation, and a hadronic representation.

The partonic representation is computed using a local \OPE at $Q^2 \ll (m_c+m_b)^2$.
The number of subtractions $u$ is chosen \emph{a posteriori} as the smallest number that guarantees convergence
of the integral on the right-hand side for the partonic representation.
One can show that, for instance, $u=1$ for $\Pi_V^0$, $u=2$ for $\Pi_V^1$, and $u=3$ for $\Pi_T^1$ (see,\eg, Refs.~\cite{Bharucha:2015bzk,Gubernari:2023puw}).
The subtraction point $Q^2$ can be chosen freely as long as it lays below all branch points of $\Pi^{J}_{\Gamma}$.

The hadronic representation can be obtained from Refs.~\cite{Okubo:1971jf,Okubo:1972ih,Boyd:1995cf}:
\begin{align}
    \label{eq:ImPi}
    \Im\,\Pi_\Gamma^{J}(t + i \eps) 
        = \frac{1}{2} 
        \P_{\mu\nu}^{J}(q)
        \sum \!\!\!\!\!\!\!\! \int\limits_H d\rho_H (2\pi)^4 \delta^{(4)}(p_H - q)
            \braket{0 | J_\Gamma^{\mu} | H(p_H)}\braket{H(p_H) | J_\Gamma^{\dagger,\nu} | 0}\Big|_{q^2=t} 
    \,,
\end{align}
where $H$ denotes a hadronic state with flavour quantum numbers $B = C = -1$. 
In the case where $H=\bar{B}^{(*)}\bar{D}^{(*)}$, the $H$-to-vacuum matrix elements can be related through crossing symmetry to the $\bar{B}^{(*)}\to D^{(*)}$ form factors. 

Global quark hadron duality implies that $\schii{J,\text{hadronic}}{\Gamma}(Q^2) = \schii{J,\text{partonic}}{\Gamma}(Q^2)$,
when accounting for all possible partonic and hadronic intermediate states.
Since the hadronic contributions are positive definite by construction, the set of hadronic states can be restricted
to a finite set, changing the equality between the partonic and hadronic contributions into an inequality,
$\schii{J,\text{hadronic}}{\Gamma}(Q^2) \leq \schii{J,\text{partonic}}{\Gamma}(Q^2)$.
This provides an upper bound for each of these form factors, which is commonly called the \emph{unitarity bound}~\cite{Okubo:1971jf,Okubo:1972ih,Boyd:1995cf}.

The unitarity bounds for the $J_V^\mu$ and $J_A^\mu$ currents in $B^{(*)}\to D^{(*)}$ decays were derived in Refs.~\cite{deRafael:1992tu,deRafael:1993ib,Boyd:1994tt,Boyd:1995cf,Boyd:1997kz,Caprini:1997mu}.
The unitarity bounds for the $J_T^\mu$ and $J_{AT}^\mu$ currents have --- to the best of our knowledge --- never been discussed in the context of $b\to c$ transitions. For rare $B\to K^{(*)}$ decays, they have been studied in Refs.~\cite{Bharucha:2010im,Gubernari:2023puw}.
In the remainder of this section, we derive for the first time the unitarity bounds for the $J_T^\mu$ and $J_{AT}^\mu$ currents in $\bar B^{(*)}\to D^{(*)}$ decays.
\\

To calculate $\schii{J}{\Gamma}$ defined in \cref{eq:chisub}, we expand the function $\Pi^{J}_{\Gamma}$ in a local \OPE for $Q^2 \ll (m_b + m_c)^2$, where the leading power term can be computed using perturbative QCD. 
Hereafter, we choose the subtraction point $Q^2=0$, as it simplifies the OPE calculation. 
Different choices of $Q^2$ only marginally impact the effectiveness of the unitarity bound~\cite{Caprini:1995wq}. 
The \OPE results up to next-to-leading order corrections in $\alpha_s$ are given in Ref.~\cite{Bharucha:2010im}.
The next-to-leading power corrections to the \OPE, i.e. the contributions of the various vacuum condensates, are negligible, due to their suppression by at least a factor of $1/m_b^5$~\cite{Boyd:1997kz,Bharucha:2010im}.
The numerical evaluation of the functions $\schii{J}{T}$ and $\schii{J}{AT}$ including $\alpha_s$ corrections yields
\begin{align}
    &
    \schi{1}{T}{\tiny \OPE}
    (Q^2=0)
    =
    4.98(40) \cdot 10^{-4} \,\GeV^{-2} \,,
    &&
    \schi{1}{AT}{\tiny\OPE}
    =
    2.77(22) \cdot 10^{-4} \,\GeV^{-2} \,.
    & 
\end{align}
These results have been calculated at the scale $\mu = \sqrt{m_c m_b} = 2.31\,\GeV$, and
we assigned an 8\% uncertainty to these central values to account for the missing $\alpha_s^2$ corrections.
\\

On the hadronic side, the first contributions to the unitarity bound arise from $b\bar{c}$ bound states, \ie, the $\bar{B}_c$, $\bar{B}_c^*$, and similar
states. These 1-body states contribute through their respective vacuum decay constants.
We do not account for these contribution in our analysis because, to the best of our knowledge, their decay constants are presently not known.\footnote{
    This is in contrast the situation for the decay constants for the axial and vector currents, which have been calculated using lattice QCD~\cite{McNeile:2012qf,Colquhoun:2015oha} and QCD sum rules~\cite{Narison:2019tym,Aliev:2019wcm}.
}
Although this weakens the statistical power of the unitarity bounds, it does not prevent us from applying them~\cite{Boyd:1995sq,Caprini:1997mu}.

The next contributions arise from 2-body states. Following the original works~\cite{Boyd:1997kz,Caprini:1997mu}, we do not account for
two-body states $H=B_c\pi$, which do not contribute in the limit of isospin symmetry.
This leads us to consider the $H=\bar{B}^{(*)}\bar{D}^{(*)}$ states next.
Using the form factor definitions given in \cref{sec:FF-def}, we express their contributions to \cref{eq:ImPi} as
\begin{align}
        \Im \Pi_{T}^{1}(t)
        \supset & \,
        \frac{n_f}{96\pi}\theta(t-t_+^{BD})
        \Bigg(
            \frac{(t-t_+^{BD})^{\frac32}(t-t_-^{BD})^{\frac32}}{t_+^{BD}}
            2|f_T|^2 
        \nonumber\\*
        + &
        4 \frac{(t-t_+^{BD^*})^{\frac32}(t-t_-^{BD^*})^{\frac32}}{t}
            |T_1|^2
        \nonumber\\*
        + &
        4 \frac{(t-t_+^{B^*D})^{\frac32}(t-t_-^{B^*D})^{\frac32}}{t}
            |\bar{T}_1|^2
        \nonumber\\*
        + & 
        4 \frac{(t-t_+^{B^*D^*})^{\frac32}(t-t_-^{B^*D^*})^{\frac32}}{t_+^{B^*D^*} - t_-^{B^*D^*}} 
        \left(
            |T_7|^2 + \frac{1}{2}|T_{10}|^2
        \right)
        \nonumber\\*
        + & 
        \frac{(t-t_+^{B^*D^*})^{\frac32}(t-t_-^{B^*D^*})^{\frac32}}{t} 
        \left(
            |T_8|^2 \!+|T_9|^2
        \right)
        \Bigg)
        \,,\\
        \Im \Pi_{AT}^{1}(t) 
        \supset & \,
        \frac{n_f}{96\pi}\theta(t-t_+^{BD})
        \nonumber\\*
        \times &
        \Bigg(
            \frac{(t-t_+^{BD^*})^{\frac12}(t-t_-^{BD^*})^{\frac12}}{t}
                \left(4 t_+^{BD^*}t_-^{BD^*} |T_2|^2 + 2 \,\frac{t}{t_+^{BD^*}}(t_+^{BD^*} - t_-^{BD^*})^2
            |T_{23}|^2   \right)
        \nonumber\\*
        + &
            \frac{(t-t_+^{B^*D})^{\frac12}(t-t_-^{B^*D})^{\frac12}}{t}
                \left(4 t_+^{B^*D}t_-^{B^*D} |\bar{T}_2|^2 + 2 \,\frac{t}{t_+^{B^*D}}(t_+^{B^*D} - t_-^{B^*D})^2
            |\bar{T}_{23}|^2   \right)
        \nonumber\\*
        + &
        4 \frac{(t-t_+^{B^*D^*})^{\frac52}(t-t_-^{B^*D^*})^{\frac12}}{t(t_+^{B^*D^*} - t_-^{B^*D^*})}
        \left(
            |T_4|^2
        \right)
        \nonumber\\*
        + &
        \frac{(t-t_+^{B^*D^*})^{\frac12}(t-t_-^{B^*D^*})^{\frac12}}{16 t}
            (t_+^{B^*D^*} - t_-^{B^*D^*})^2
            \left(
                |T_5|^2+|T_6|^2
            \right)
        \Bigg)
        \,.
\end{align}
with $t_\pm^{B^{(*)}D^{(*)}} \equiv (m_{B^{(*)}} \pm m_{D^{(*)}})^2$.
Here $n_f=2$ for $\bar{B}^{(*)}\to D^{(*)}$ processes, due to the assumed isospin symmetry,
and $n_f=1$ for $\bar{B}_s^{(*)}\to D_s^{(*)}$ processes.
We assume that the pair-production threshold and branch points for all processes considered in this article are the same, i.e., $\theta(t - t_+^{BD})$.
This follows the approach of the original works~\cite{Boyd:1997kz,Caprini:1997mu} and is justified by the relatively small differences in masses.
For a more general treatment beyond this approximation, see Refs.~\cite{Boyd:1995sq,Caprini:1995wq,Gubernari:2020eft,Blake:2022vfl,Flynn:2023qmi,Gubernari:2023puw,Gopal:2024mgb}.

The contribution to the imaginary part of $\Pi_{\Gamma}^{J}$ due to any generic \BBstarToDDstar form factor $F$ can then be expressed as
\begin{equation}
\begin{aligned}
    \Im \Pi_{\Gamma}^{J}(t) & >
    \frac{n_f}{K L\pi}
    (t - t_+)^{\frac{a}{2}}
    (t - t_-)^{\frac{b}{2}}
    t^{-(c-u+2)}
    |F(t)|^2
    \theta(t-t_+)
    \\
    & \equiv 
    W_F(t)
    |F(t)|^2
    \theta(t-t_+)
    \,,
    \label{eq:weightfun}
\end{aligned}
\end{equation}
where in the second line we have defined the weight function $W_F$.
Note that \cref{eq:weightfun} is a more general version of eq.~(4.13) of Ref.~\cite{Boyd:1997kz}.
Here we introduce the additional parameter $u$, which as discussed above is the minimal number of subtractions needed. 
The values of the parameters $K,\,L,\,a,\,b,$ and $c$ for each tensor form factor are listed in \cref{tab:outer}.
The values of these parameters for the (axial-)vector $B^{(*)}\to D^{(*)}$ form factors can be extracted from Refs.~\cite{Caprini:1997mu,Boyd:1997kz}.
To proceed, we define the map
\begin{align}
    z(q^2) = 
    \frac{\sqrt{t_+ - q^2} - \sqrt{t_+ - t_0^{\phantom{2}}}}{\sqrt{t_+ - q^2} + \sqrt{t_+ - t_0^{\phantom{2}}}}
    \,,
    \label{eq:zmap}
\end{align}
where the parameter $t_0$ can be arbitrarily chosen in the interval  $(-\infty,t_+)$.
In this article, we choose $t_0=t_-$, as in Refs.~\cite{Caprini:1997mu,Bordone:2019guc}.
Following Ref.~\cite{Caprini:1997mu} we rewrite \cref{eq:zmap} in terms of the kinematical variable $w$ defined in \cref{eq:wdef}:
\begin{align}
    \label{eq:ztow}
    z(w) = \frac{\sqrt{1+w} - \sqrt{2}}{\sqrt{1+w} + \sqrt{2}} \,.
\end{align}
The variable $w$ is convenient, because the various $B^{(*)}D^{(*)}$ thresholds --- which occur for different values of $t$ --- all occur at $w=-1$.
Equation (\ref{eq:zmap}) maps the complex $q^2$ plane on the unit disk of the complex $z$ plane.
Furthermore, we introduce the \emph{outer functions}, which are defined such that they fulfil the equation
\begin{align}
    &
    |\phi_F (t)|^2 = 
    \left.
    \frac{
        W_F(t)
        }{
        \left| \frac{dz(\theta)}{d\theta}
        \frac{dt(z)}{dz} \right|
        t^{(u+1)}
        \schii{\lambda}{\Gamma}(0)
        }
    \right|_{\scriptsize \begin{array}{c}t=t(z)\\[-1mm] z=e^{i\theta}\end{array}}
    &
    & 
    \text{ for }|z| = 1 
    \,,
    \label{eq:defouter}
\end{align}
and they are analytical on the open unit disk, that is for $|z| < 1$.
We use the formula for the outer functions given in Ref.~\cite{Boyd:1997kz}, which reads
\begin{eqnarray}
    \label{eq:outert}
    \phi_F (t) & = & \sqrt{\frac{n_f}{K \pi \chi}} \,
    \left( \frac{t_+ - t}{t_+ - t_0} \right)^{\frac 1 4}
    \left( \sqrt{t_+ - t} + \sqrt{t_+ - t_0} \right)
    \left( t_+ - t \right)^{\frac a 4} \nonumber \\* & &
    \times \left( \sqrt{t_+ - t} + \sqrt{t_+ - t_-} \right)^{\frac b 2}
    \left( \sqrt{t_+ - t} + \sqrt{t_+} \right)^{-(c+3)} .
\end{eqnarray}

\begin{table}[t!]
    \newcommand{\pp}{\phantom{+}}
    \centering
    \renewcommand{\arraystretch}{1.2}
    \begin{tabular}{l @{\hspace{2.5em}} c @{\hspace{2.5em}} cccccc}
        \toprule
        \multirow{2}{*}{Process}        &
        \multirow{2}{*}{Form Factor}    &
        \multicolumn{5}{c}{Parameters}  \\
                                        & 
                                        & 
        $K$                             &
        $\chi$                          &
        $a$                             &
        $b$                             &
        $c$                             \\
        \toprule       
        \BToD                           & 
        $f_T$                           & 
        $48 \, t_+$                     &
        $\schii{1}{T}$                  &
        3                               &
        3                               &
        1                               \\
        \midrule
        \multirow{3}{*}{\BToDstar}      & 
        $T_1$                           & 
        $24$                            &
        $\schii{1}{T}$                  &
        3                               &
        3                               &
        2                               \\
                                        & 
        $T_2$                           & 
        $24/(t_+ t_-)$                  &
        $\schii{1}{AT}$                 &
        1                               &
        1                               &
        2                               \\
                                        & 
        $T_{23}$                        & 
        $48\,t_+/(t_+ - t_-)^2$         &
        $\schii{1}{AT}$                 &
        1                               &
        1                               &
        1                               \\
        \midrule
        \multirow{3}{*}{\BstarToD}      & 
        $\bar{T}_1$                     & 
        $24$                            &
        $\schii{1}{T}$                  &
        3                               &
        3                               &
        2                               \\
                                        & 
        $\bar{T}_2$                     & 
        $24/(t_+ t_-)$                  &
        $\schii{1}{AT}$                 &
        1                               &
        1                               &
        2                               \\
                                        & 
        $\bar{T}_{23}$                  & 
        $48\,t_+/(t_+ - t_-)^2$         &
        $\schii{1}{AT}$                 &
        1                               &
        1                               &
        1                               \\
        \midrule
        \multirow{5}{*}{\BstarToDstar}  & 
        $T_4$                           & 
        $24 \, (t_+ - t_-)$             &
        $\schii{1}{AT}$                 &
        5                               &
        1                               &
        1                               \\
                                        & 
        $T_5,\,T_6$                     & 
        $2^9\cdot3/(t_+ - t_-)^2$       &
        $\schii{1}{AT}$                 &
        1                               &
        1                               &
        2                               \\
                                        & 
        $T_7$                           & 
        $24\, (t_+ - t_-)$              &
        $\schii{1}{T}$                  &
        3                               &
        3                               &
        1                               \\
                                        & 
        $T_8,\,T_9$                     & 
        $96$                            &
        $\schii{1}{T}$                  &
        3                               &
        3                               &
        2                               \\
                                        & 
        $T_{10}$                        & 
        $48\, (t_+ - t_-)$              &
        $\schii{1}{T}$                  &
        3                               &
        3                               &
        1                               \\
        \bottomrule
    \end{tabular}
    \caption{
    \label{tab:outer} 
      Parameters of the weight functions in \cref{eq:weightfun}. 
      The number of subtractions $u=3$ is the same for all tensor form factors.
    }
\end{table}

Using \cref{eq:zmap,eq:ztow,eq:defouter}, the unitarity bound (\ref{eq:chisub}) can be written as 
\begin{align}
    \left.
    1
    > \frac{1}{2\pi}
    \int\limits_{-\pi}^{+\pi}
    d\theta\,\sum_F \left|\phi_F (t) F(t)\right|^2
    \,\,\right|_{\scriptsize \begin{array}{c}t=t(z)\\[-1mm] z=e^{i\theta}\end{array}}
    \,,
    \label{eq:bound1}
\end{align}
where the sum runs over all the form factors contributing to a given spin parity channel.
The form factors may present poles below the $t_+$ threshold, which can be removed by including the \emph{Blaschke factors}
\begin{align}
    P_{J^P}(t) \equiv 
    \prod_j
    \frac{z(t) - z(m_{J^P,j}^2)}{1 - z(t) \, z^*(m_{J^P,j}^2)}
    \,.
\end{align}
Here, $m_{J^P,j}$ are the masses of the $c\bar{b}$ bound states below the threshold that contribute to the function $\Pi_{J^P}^{(T)}$.
These masses are listed in Table III of Ref.~\cite{Bigi:2017jbd}, incorporating the updates mentioned in Footnote 3 of Ref.~\cite{Gambino:2019sif}.
For the reader's convenience, these results are summarized in \cref{tab:Blaschke}.
Therefore, \cref{eq:bound1} becomes
\begin{align}
    \left.
    1 > \frac{1}{2\pi}
    \int\limits_{-\pi}^{+\pi}
    d\theta\,\sum_F \left|P_{J^P}(t) \phi_F (t) F(t)\right|^2
    \,\,\right|_{\scriptsize \begin{array}{c}t=t(z)\\z=e^{i\theta}\end{array}}
    \,,
    \label{eq:bound2} 
\end{align}
since $|P_{J^P}|^2=1$ for $|z|=1$.
The function inside the modulus squared in \cref{eq:bound2} is now analytic on the open unit disk and hence can be expanded in a Taylor series,
\begin{align}
    P_{J^P}(t) \phi_F (t) F(t) = \sum_{n=0}^\infty a_n^F z^n(t)
    \qquad \implies \qquad
    F(t) = \frac{1}{P_{J^P}(t) \phi_F (t)} \sum_{n=0}^\infty a_n^F z^n(t)
    \,.
    \label{eq:taylor}
\end{align}
Finally, substituting \cref{eq:taylor} into \cref{eq:bound2}, we obtain the unitarity bound in a particularly convenient form, i.e. an explicit constraint on the coefficients of the $z$ expansion:
\begin{align}
    \label{eq:bound-z}
    \sum_F \sum_{n=0}^\infty |a_n^F|^2 < 1
    \,.
\end{align}
There is one of these constraints for each spin-parity channel.
Clearly, the greater the number of form factors considered simultaneously in a fit, the more constraining \cref{eq:bound-z} becomes.
In addition to the \BstarToDstar and \BsstarToDsstar form factors, this framework can also incorporate, e.g., the $\bar{B} \to D^{**}$ and $\Lambda_b \to \Lambda_c$ form factors, which have been theoretically predicted~\cite{Detmold:2015aaa,Gubernari:2022hrq,Gubernari:2023rfu}.

\begin{table}[t!]
    \newcommand{\pp}{\phantom{+}}
    \centering
    \renewcommand{\arraystretch}{1.2}
    \resizebox{1.0\textwidth}{!}{%
    \begin{tabular}{l @{\hspace{1em}} cc @{\hspace{1em}} ccc @{\hspace{1em}} ccc @{\hspace{1em}} cccc}
        \toprule
        $J^P$ &
        \multicolumn{2}{c}{$0^+$} &
        \multicolumn{3}{c}{$0^-$} &
        \multicolumn{3}{c}{$1^-$} &
        \multicolumn{4}{c}{$1^+$} \\
        $j$ &
        1 & 2 &         
        1 & 2 & 3 &     
        1 & 2 & 3 &     
        1 & 2 & 3 & 4 \\
        \midrule
        Mass [GeV] &
        6.704 &
        7.122 &
        6.275 &
        6.871 &
        7.250 &
        6.329 &
        6.910 &
        7.020 &
        6.739 &
        6.750 &
        7.145 &
        7.150 \\
        \midrule
        Ref.  &
        \cite{PDG:2024cfk,Dowdall:2012ab}  &
        \cite{Godfrey:2004ya} &
        
        \cite{PDG:2024cfk,McNeile:2012qf} &
        \cite{Sirunyan:2019osb,Aaij:2019ldo} &
        \cite{Godfrey:2004ya} &
        
        \cite{PDG:2024cfk,Dowdall:2012ab} &
        \cite{Dowdall:2012ab} &
        \cite{Rai:2014fga} &

        \cite{Dowdall:2012ab} &
        \cite{Godfrey:2004ya} &
        \cite{Godfrey:2004ya} &
        \cite{Godfrey:2004ya} \\
        \bottomrule
    \end{tabular}
    }
    \caption{ 
    \label{tab:Blaschke} 
      Masses of the $\bar{c} b$ QCD bound states organised according to their spin $J$ and parity $P$ quantum numbers.
    }
\end{table}


\section{Analysis setup} 
\label{sec:setup}

Our analyses are conducted by using the \EOS software~\cite{EOSAuthors:2021xpv} in version 1.0.15~\cite{EOS:v1.0.15}. The contents of this section provide a detailed explanation of
the analysis file used in the course of this study. The analysis file is available
as part of the supplementary material~\cite{EOS-DATA-2025-02}.

\subsection{Statistical framework}
\label{sec:setup:framework}

Our study is based on a Bayesian analysis of the theoretical data for \BqToDqDqStar form factors.
A central element to all Bayesian analyses is the posterior probability density function (posterior PDF or just posterior).
It is defined as
\begin{equation}
    \label{eq:pheno:setup:posterior}
    P(\vecx\,|\, D, M) \equiv \frac{P(D \,|\, \vecx, M)\, P_0(\vecx \,|\, M)}{P(D \,|\, M)}\,,
\end{equation}
where $\vec{x}$ represents the parameters, $D$ represents a data set, and $M$ represents a fit model.
Statistical information about the form factors obtained by theory groups is encoded in the likelihood
$P(D \,|\, \vecx, M)$. We discuss the individual data sets that enter the likelihood in \cref{sec:setup:data}.
As part of the likelihood, we impose the unitarity bounds on the HQE parameter space; details are provided in \cref{sec:setup:unitarity-bounds}.
Information about the form factors obtained prior or independently of our analysis, as well as information inherent to the
fit model, is represented by the prior PDF $P_0$. We discuss the applicable fit models in \cref{sec:setup:models}.
The evidence $P(D \,|\, M) \equiv \int d\vecx P(D \,|\, \vecx, M) P_0(\vecx \,|\, M)$ ensures that the posterior is genuinely a PDF, \ie, it is
normalized to unity when integrated over the parameters.

In the course of our analysis, we both maximize this posterior PDF with respect to $\vecx$ (providing us with information
about the best description of the data within a given model)
and draw random samples $\vecx \sim P(\vecx\,|\, D,M)$ (thereby determining the allowed parameter space and producing posterior predictions).
The random samples, together with their associated posterior values, further permit us to
compute the evidence $P(D \,|\, M)$ associated with a fit model $M$.
To draw the posterior samples, \EOS uses dynamical nested sampling~\cite{Higson:2018}
as implemented in the open-source \texttt{dynesty} software~\cite{dynesty:v2.0.3}.
The use of dynamical nested sampling ensures simultaneously high accuracy for the posterior predictions
and the estimate of the evidence.
\\

Using the evidence estimates, we can compare pairs of fit models $M_1$ and $M_2$ using their Bayes factor
\begin{equation}
    K(M_1, M_2) \equiv P(D \,|\, M_1) / P(D \,|\, M_2)\,.
\end{equation}
The Bayes factor provides information about the relative efficiency of the two models in describing their common dataset $D$.
Following the textbook by Jeffreys~\cite{Jeffreys:1939xee}, the model $M_1$ is preferred over the model $M_2$ if $K(M_1, M_2) > 1$.
The size of the Bayes factor provides information about the level of this preference, which ranges
from ``barely worth mentioning'' ($1 < K(M_1, M_2) \leq 3$) via substantial ($3 < K(M_1, M_2) \leq 10$)
and  strong ($10 < K(M_1, M_2) \leq 100$) to decisive ($100 < K(M_1, M_2)$).
If $K(M_1, M_2) < 1$, the above interpretation applies when exchanging $M_1$ and $M_2$.

\subsection{Theoretical data sets}
\label{sec:setup:data}

All of our statistical analyses involve one or more of the following theoretical data sets.
Each data set provides a (multivariate) Gaussian contribution to the overall likelihood.
\begin{description}
    \item[{\FNALMILC[~2012]}] The \FNALMILC collaboration has published results for the form factor ratio $f_0^{\BsToDs}(m_\pi^2)/f_0^{\BToD}(m_\pi^2)$~\cite{Bailey:2012rr}.
    The total number of observations is $1$.

    \item[\texttt{ABMS 2012}\xspace] In Ref.~\cite{Atoui:2013zza} the authors have published the ratio $f_T/f_+$ at a single point $q^2 = 11.5 \, \GeV^2$
    for both \BToD and \BsToDs transitions.
    The total number of observations is $2$.
    
    \item[{\FNALMILC[~2015]}] The \FNALMILC collaboration has published results for the \BToD form factors $f_+$ and $f_0$~\cite{MILC:2015uhg}.
    We generate and use synthetic data points derived from these results at four different values of the momentum transfer:
    \begin{equation}
        q^2 \in \lbrace 0, 8.49\,\GeV^2, 10.07\,\GeV^2, 11.64\,\GeV^2\rbrace\,.
    \end{equation}
    The form factors $f_+$ and $f_0$ fulfil an equation of motion constraint at $q^2 = 0$, $f_+(0) = f_0(0)$,
    which reduces the number of observations by $1$.
    The total number of observations is $7$.

    \item[{\HPQCD[~2015]}] The \HPQCD collaboration has published results for the \BToD form factors $f_+$ and $f_0$~\cite{Na:2015kha}.
    We generate and use synthetic data points derived from these results at three different values of the momentum transfer:
    \begin{equation}
        q^2 \in \lbrace 0, 9.28\,\GeV^2, 11.64\,\GeV^2\rbrace\,.
    \end{equation}
    As for \FNALMILC 2015, an equation of motion reduces the number of observations by $1$.
    The total number of observations is $5$.

    \item[{\HPQCD[~2019]}] The \HPQCD collaboration has published results for the \BsToDs form factors $f_+$ and $f_0$~\cite{McLean:2019qcx}.
    We generate and use synthetic data points from these results at three different values of the momentum transfer:
    \begin{equation}
        q^2 \in \lbrace 
        0, 
        5.78\,\GeV^2, 
        11.54\,\GeV^2
        \rbrace\,.
    \end{equation}
    As for \FNALMILC 2015, an equation of motion reduces the number of observations by $1$.
    The total number of observations is $5$.

    \item[{\FNALMILC[~2021]}] The \FNALMILC collaboration has published results for the four \BToDstar form factors needed
    for predictions within the SM~\cite{FermilabLattice:2021cdg}. 
    We convert the three data points for each form factor given in the ancillary files of this article into the basis $V$, $A_0$, $A_1$, and $A_{12}$.
    The total number of observations is $12$.
    
    \item[{\HPQCD[~2023]}] The \HPQCD collaboration has published results for all $14$ \BToDstar and \BsToDsstar form factors~\cite{Harrison:2023dzh}.
    The authors provide synthetic data points at five different $q^2$ points.
    To prevent numerical instability in the inversion of the covariance matrix, we discard the second-lowest \( q^2 \) point while retaining the \( q^2 = 0 \) point.
    The total number of observations is $56$.
    
    \item[{\JLQCD[~2023]}] Similarly to the \FNALMILC collaboration, the \JLQCD  collaboration has published results for the four \BToDstar form factors needed for SM predictions only~\cite{Aoki:2023qpa}.
    We convert the three data points for each form factor given in Ref.~\cite{Aoki:2023qpa} into the basis $V$, $A_0$, $A_1$, and $A_{12}$.
    The total number of observations is $12$.

    \item[{\LQCD}] This label stands in for the use of all of the above data sets, which have all been obtained from lattice QCD simulations.
    The total number of observations is 100.
    
    \item[{\QCDSR}] We further use a collection of QCD sum rule results as compiled (and partially re-derived) in our previous analysis~\cite{Bordone:2019guc}.
    These results are obtained from three-point and four-point QCD sum rules for the Isgur-Wise parameters $\chi_{2q}(1)$, $\chi_{2q}^{(1)}(1)$, $\chi_{3q}^{(1)}(1)$, $\eta_{q}(1)$, $\eta_{q}^{(1)}(1)$
    \cite{Neubert:1992wq,Neubert:1992pn,Ligeti:1993hw}
    and QCD light-cone sum rules for all \BToD, \BsToDs, \BToDstar and \BsToDsstar form factors~\cite{Gubernari:2018wyi,Bordone:2019guc}, with the exception of the $\BToD$ and $\BsToDs$ tensor
    form factors.
    The total number of observations is $76$.
\end{description}
Correlations between the above-listed data sets are not available and we make no attempt to take them into account.
Correlations of the observations within a given data set are used, if they have been made available.

Data sets involving form factors of the tensor currents $\bar{c} \gamma^{\mu\nu} (\gamma_5)b$ require additional treatment,
since these currents need to be renormalised and are therefore scale-dependent quantities.
Hence, wherever form factors of the tensor current are used,
this scale dependence has to be taken into account. In order to obtain a given matrix element, given at a scale $\mu_1 \leq m_b$, at a scale $\mu_2 < \mu_1$, we write schematically
\begin{equation}
    \braket{\bar{c} \sigma^{\mu\nu} b}_{\mu_2} = U(\mu_2, \mu_1) \braket{\bar{c} \sigma^{\mu\nu} b}_{\mu_1}\,,
\end{equation}
with the evolution factor $U(\mu_2,\mu_1)$ cancelling the corresponding factor from the evolution of the Wilson coefficients,
\begin{equation}
    C_{\mu_2}\braket{\bar{c} \sigma^{\mu\nu} b}_{\mu_2} = C_{\mu_1} \braket{\bar{c} \sigma^{\mu\nu} b}_{\mu_1}\,.
\end{equation}
Writing in leading-logarithmic approximation
\begin{equation}
    \frac{dC_T}{d\ln\mu} = \hat\gamma_T C_T = \frac{\alpha_s}{4\pi}\tilde \gamma_T C_T\,,
\end{equation}
where the anomalous dimension $\tilde\gamma_T=8/3$ can be taken from Ref.~\cite{Aebischer:2017gaw}, we obtain at this order the standard result
\begin{equation}\label{eq:evolution}
    U(\mu_2,\mu_1) = \left(\frac{\alpha_s(\mu_2)}{\alpha_s(\mu_1)}\right)^{+\frac{\tilde\gamma_T}{2\beta_0}}
    \qquad
    \text{with }
    \beta_0 \equiv \beta_0^{n_f = 4} = \frac{25}{3}
    \,.
\end{equation}
Given the proportionality between form factors and matrix elements, with scale-independent coefficients, we therefore obtain the universal scaling 
\begin{equation}
    h_{T_{(i)}}(\mu_2) = U(\mu_2,\mu_1)h_{T_{(i)}}(\mu_1)\,,
\end{equation}
with the evolution factor given in Eq.~\eqref{eq:evolution}.
We choose to renormalize the
theory predictions for the tensor form factors at the matching scale $\mu_2^2= m_b m_c$,
as is usual for HQET matching calculations~\cite{Falk:1992wt,Bernlochner:2017jka}.
Hence, lattice data that is renormalized at \eg the scale $\mu_1 = 4.8\,\GeV$, as is the case for the \HPQCD[~2023] data set, will be multiplied by a factor $U(\sqrt{m_bm_c}, 4.8\,\GeV) \simeq 1.03$ before being compared to the \HQE theory predictions for the form factors.

\subsection{Fit models and parameters}
\label{sec:setup:models}

Working within the framework of the \HQE as introduced in \cref{sec:theory:hqe}, we parametrize the various
\BBstarToDDstar form factors through a systematic expansion of the relevant Isgur-Wise functions.
Following Refs.~\cite{Jung:2018lfu,Bordone:2019vic,Bordone:2019guc}, we start with a Taylor expansion of the LP
Isgur-Wise functions $\xi_q(w)$ about the point $w = 1$,
\begin{equation}
    \label{eq:LPexp}
    \xi_q(w) = 1 + \xi_q^{(1)}(1) (w - 1) + \frac{1}{2} \xi_q^{(2)}(1) (w - 1)^2 + \frac{1}{6} \xi_q^{(3)}(1) (w - 1)^3\,,
\end{equation}
where $q=u/d,s$ represents the spectator quark flavour.
We then further expand each of the monomials of $(w - 1)$
in terms of $z$. In this second expansion, we ensure that only terms up to order $N_\text{LP} \leq 3$ are kept; all terms
of higher orders in $z$ are discarded.\footnote{%
    For $t_0=t_-$ a term $(w - 1)^n$ only produces terms of order $z^n$ or higher orders.
}
The coefficients $\xi_q^{(n)}(1)$ represent our choice of fit parameters at leading-power.
\\
At NLP, three additional Isgur-Wise functions enter for each of the two spectator quark flavours.
These functions are labelled $\chi_{2q}(w)$, $\chi_{3q}(w)$, and $\eta_q(w)$. Using $\eta_q$ as an example,
we expand these functions as
\begin{equation}
    \label{eq:NLPexp}
    \eta_q(w) = \xi_q(w) \hat\eta_q(w) = \xi_q(w) \left[\hat\eta_q(1) + \hat\eta_q^{(1)}(1) (w - 1) + \frac{1}{2} \hat\eta_q^{(2)}(1) (w - 1)^2\right]\,.
\end{equation}
We expand each monomial in $w - 1$ in terms of $z$ and keep only terms up to order $N_\text{NLP}$.
The normalisation to the leading-power function $\xi_q(w)$ ensures that we only need to parametrize the deviation
from the leading-power behaviour. This motivates our decision to impose $N_\text{NLP} < N_\text{LP}$ for all
of our fit models.
The coefficient $\hat\chi_{3q}(1)$ is set to zero manifestly as a consequence of Luke's theorem \cite{Luke:1990eg}; all other coefficients represent our choice
of fit parameters at next-to-leading power.
\\
For \BBstarToDDstar form factors, six additional functions enter for each of the two spectator quark flavours
at NNLP in $\varepsilon_c$~\cite{Falk:1992wt}. Following the notation of Ref.~\cite{Bordone:2019vic}, we label
these functions $\ell_{1q}(w)$ through $\ell_{6q}(w)$,
expanding them as
\begin{equation}
    \label{eq:NNLPexp}
    \ell_{iq}(w) = \xi_q(w) \hat\ell_{iq}(w) = \xi_q(w) \left[\hat\ell_{iq}(1) + \hat\ell_{iq}^{(1)}(1) (w - 1)\right]\,,\,\, i\in[1,6]\,,\,q\in\{d,s\}\,.
\end{equation}
We again expand each monomial in $w - 1$ in terms of $z$ and keep only terms up to order $N_\text{NNLP} < N_\text{NLP}$.
The Taylor coefficients represent our choice of fit parameters at next-to-next-to-leading power.\\

Using the parametrisation introduced above to fit the data sets listed in \cref{sec:setup:data} other than the \QCDSR data set
leads to up to four blind directions in the posterior.
This is a consequence of the qualitatively different constraints
obtained in the \QCDSR data set and the other data sets. Specifically, in the \QCDSR data, we have direct control of
the NLP Isgur-Wise function parameters at $w=1$, through constraints on $\hat\chi_{2q}(1)$, $\hat\chi_{2q}^{(1)}(1)$, $\hat\chi_{3q}^{(1)}(1)$, $\hat\eta_{q}(1)$, and $\hat\eta_{q}^{(1)}(1)$.
In contrast, all other data sets constrain only the full \BToDDstar form factors, which correspond to
fixed linear combinations of the various Isgur-Wise functions.
When fitting to only the latter data sets, we choose to remove the blind directions from our analysis
by applying the following redefinitions of the fit parameters in a subset of our fit models
(in order, from top to bottom and left to right):
\begin{equation}
\begin{aligned}
    \hat\chi_{2q}(1) 
        & = \tilde\chi_{2q}(1)       - \frac{1}{4} \varepsilon_c \hat\ell_{3q}(1)       \,, &
    \hat\chi_{2q}^{(1)}(1)
        & = \tilde\chi_{2q}^{(1)}(1)       - \frac{1}{4} \varepsilon_c \hat\ell_{3q}^{(1)}(1)       \,,\\
    \hat\chi_{3q}^{(1)}(1)
        & = \tilde\chi_{3q}^{(1)}(1) - \frac{1}{12}\varepsilon_c \tilde\ell_{1q}^{(1)}(1) \,, \\
    \hat\ell_{1q}^{(1)}(1)
        & = \tilde\ell_{1q}^{(1)}(1) -                        \hat\ell_{3q}(1)       \,, &
    \hat\ell_{2q}^{(1)}(1)
        & = \tilde\ell_{2q}^{(1)}(1) - \frac{1}{3}            \tilde\ell_{1q}^{(1)}(1) \,.&&\\
\end{aligned}
\end{equation}
These redefinitions are equivalent to setting $\hat\ell_{1q}^{(1)}(1)$, $\hat\ell_{3q}(1)$, and $\hat\ell_{3q}^{(1)}(1)$ to zero.
They leave all likelihoods except for the \QCDSR one invariant, up to corrections of higher powers in $\varepsilon_Q$ and $z$
than those considered in this study. As a consequence, we remove the parameters $\hat\ell_{1q}^{(1)}(1)$, $\hat\ell_{3q}(1)$, and $\hat\ell_{3q}^{(1)}(1)$ from the set of free fit parameters for fits that exclude the \QCDSR data set.
\\

Based on the above, we define the following fit models:
\begin{description}
    \item[$\boldsymbol{2/1/0^*}$] This model is characterized by $N_\text{LP} = 2$, $N_\text{NLP} = 1$, and $N_\text{NNLP} = 0$.
    All parameters for $q=u/d$ and $q=s$ spectator quarks are treated as independent. Two of the NNLP parameters, \ie $\hat\ell_{3d}(1)$ and $\hat\ell_{3s}(1)$, are removed to avoid blind directions.
    This yields a total number of $24$ free fit parameters.

    \item[$\boldsymbol{2/1/0^*}$ w/ \SUF] This model is obtained from the $2/1/0^*$ model by identifying the parameters $\hat\ell_{jd}(1)$ and $\hat\ell_{js}(1)$ with each other. 
    This reduces the total number of free fit parameters to $19$.
    
    \item[$\boldsymbol{3/2/1^*}$] This model is characterized by $N_\text{LP} = 3$, $N_\text{NLP} = 2$, and $N_\text{NNLP} = 1$.
    All parameters for $q=u/d$ and $q=s$ spectator quarks are treated as independent. For each spectator flavour, three of the NNLP parameters are removed to avoid blind directions.
    This yields a total number of $40$ free fit parameters.

    \item[$\boldsymbol{3/2/1^*}$ w/ \SUF] This model is obtained from the $3/2/1^*$ model by identifying the parameters $\hat\ell_{jd}^{(n)}(1)$ and $\hat\ell_{js}^{(n)}(1)$ with each other. 
    This reduces the total number of free fit parameters to $31$.

    \item[$\boldsymbol{3/2/1}$] This model is characterized by $N_\text{LP} = 3$, $N_\text{NLP} = 2$, and $N_\text{NNLP} = 1$.
    All parameters for $q=u/d$ and $q=s$ spectator quarks are treated as independent. 
    This yields a total number of $46$ free fit parameters.
\end{description}

For all of these models, we use uniform prior distributions for the hadronic parameters,
chosen large enough so that the posterior distribution is not affected but narrow enough
to ensure efficient sampling from the posterior.
The intervals for these uniform distributions are documented as part of the
supplementary material~\cite{EOS-DATA-2025-02} within the analysis file.

\subsection{Implementation of the unitarity bounds}
\label{sec:setup:unitarity-bounds}

The simultaneous application of the HQE and unitarity bounds~\cite{Boyd:1995cf,Caprini:1997mu}
leads to exploitable approximate relations between the expansion parameters for the SM form factors.
These relations gave rise to the successful CLN parametrization~\cite{Caprini:1997mu}, which has been used ubiquitously
in past experimental and phenomenological analyses. However,
the approximations used in this parametrization are no longer appropriate given present experimental data sets with their large statistical power~\cite{Bigi:2016mdz,Jung:2018lfu,Bordone:2019vic}.
A substantial improvement to this parametrization was suggested in Ref.~\cite{Bernlochner:2017jka},
which includes both radiative and leading power corrections to the HQE relation; however, the unitarity bounds
are not accounted for in this improved approach. They have been updated in Refs.~\cite{Bigi:2016mdz,Bigi:2017jbd,Grinstein:2017nlq}.
\\

Here, we follow a different approach to employ the unitarity bounds~\cite{Bordone:2019vic,Bordone:2019guc}.
For a given form factor $F$ we match its BGL parametrization with its expression in the HQE.
Accounting for the BGL outer function $\phi_F$ and the Blaschke factor $P_F$, we determine the $n$th-order BGL coefficient as
\begin{equation}
    \label{eq:aHQE}
    a_n^F\bigg|_\text{HQE}
        = \frac{1}{n!} \frac{\partial^n}{\partial z^n} \left[ \phi_F P_F F^\text{HQE}\right]\bigg|_{z = 0}\,.
\end{equation}
Here, $F^\text{HQE}$ is the HQE representation of the form factor.
When computing the saturation of the unitarity bounds, we account for coefficients up to order $N_\text{UB}$.

In the case of the (axial-)tensor form factors this can be achieved by using the definitions
in \cref{eq:BtoD-trad1,eq:BtoDst-trad1,eq:BtoDst-trad2,eq:BtoDst-trad3,eq:BsttoD-trad1,eq:BsttoD-trad2,eq:BsttoD-trad3,eq:BstToDst-trad1,eq:BstToDst-trad2},
the inverse of the relations in \cref{app:form-factor-decomposition}, the power expansion of the HQE form factors in
\cref{eq:hTexp,eq:hT1exp,eq:hT2exp,eq:hT3exp,eq:hT1bexp,eq:hT2bexp,eq:hT3bexp,eq:hT4exp,eq:hT5exp,eq:hT6exp,eq:hT7exp,eq:hT8exp,eq:hT9exp,eq:hT10exp},
and finally \cref{eq:LPexp,eq:NLPexp,eq:NNLPexp}.
For the remaining form factors, this has been discussed in Refs.~\cite{Bordone:2019guc,Bordone:2019vic}.

Clearly, the \HQE determinations of the BGL coefficients $a_n^F\big|_\text{HQE}$ depend on multiple choices, including the truncation orders
of the $\alpha_s$ expansion, the power expansion in $\varepsilon_Q$, and
the Taylor expansion in $z$ for the Isgur-Wise function parametrization as discussed in \cref{sec:setup:models}.
When the truncation order in $z$ is chosen smaller or equal to $N_\text{UB}$, there is a risk
to misestimate the level of saturation, in particular to estimate saturation larger than the true value~\cite{Bordone:2019vic,Bordone:2019guc}.
Hence, in all analyses presented here, we use $N_\text{UB} = N_{\text{LP}} - 1$.
In our analysis, we work strictly to NLO in $\alpha_s$, NLP in $\varepsilon_b$, and NNLP in $\varepsilon_c$, which governs
the accuracy to which we can compute the BGL coefficients.
We use the unitarity bounds in the \emph{strong} formulation~\cite{Boyd:1995cf,Boyd:1997kz,Bigi:2016mdz,Bigi:2017jbd}, \ie,
\begin{equation}
    \text{Saturation}_{J^P \Gamma} = \sum_F \sum_{n=0}^{N_\text{UB}} |a_n^F|^2\,,
\end{equation}
for each of the six combination $J^P\Gamma \in \lbrace 0^+ V,\,0^- A,\,1^- V,\,1^+ A,\,1^- T,\,1^+ AT\rbrace$,
and we sum over all \BBstarToDDstar form factors $F$ that match the current $\Gamma$, total angular momentum $J$ and parity $P$.
Accounting in this way for \BBstarToDDstar form factors has the potential to strengthen the bounds beyond the traditional approach, since
the \HQE is more predictive: it predicts \emph{all} of the \BBstarToDDstar form factors.
Including all predicted form factors in the bounds is expected to lead to a greater level of saturation than in the BGL approach,
thereby constraining the \HQE parameter space more effectively.
The truncation order $N_\text{UB}$ is chosen to be $1$ for the $2/1/0^{(\ast)}$-type models
and $2$ for the $3/2/1^{(\ast)}$-type models.
\\

To implement the bounds into our analysis, we include a penalty function for each of the six bounds as part of the
likelihood~\cite{Bordone:2019guc}:
\begin{equation}
    -2 \ln \text{Penalty}_{J^P \Gamma} = \begin{cases}
        0
            & \text{if}\quad\text{Saturation}_{J^P \Gamma} \leq 1\\
        \dfrac{\left(1 - \text{Saturation}_{J^P \Gamma}\right)^2}{\sigma_{J^P \Gamma}^2}
            & \text{otherwise}
    \end{cases}\,.
\end{equation}
Here $\sigma_{J^P\Gamma}$ is chosen to reflect the relative theory uncertainty on the OPE
calculation of the quantity $\chi_\Gamma^J$.

\section{Results} 
\label{sec:results}

\subsection{Summary of the fit results for all models}
\label{sec:results:summary}

\begin{table}[t!]
    \newcommand{\pp}{\phantom{+}}
    \centering
    \renewcommand{\arraystretch}{1.30}
    \begin{tabular}{ccccccc}
        \toprule
        \multicolumn{2}{c}{Fit Model}        
            \\
        Truncation                 &
            \SUF                   &
            Data sets              &
            $\chi^2$               &
            d.o.f.                 &
            $p$-value [\%]         &
            $\ln P(D\,|\,M)$       \\
        \midrule
            \multirow{2}{*}{2/1/0$^*$}&
            \checkmark             &
            \LQCD                  & 
            52.79                  &
            81                     &
            99.36                  &
            277.9                  \\
                                   & 
            ---                    &
            \LQCD                  &
            42.10                  &
            76                     &
            99.94                  &
            269.7                  \\
        \midrule
            \multirow{2}{*}{3/2/1$^*$}& 
            \checkmark             & 
            \LQCD                  &
            43.33                  &
            69                     &
            99.34                  &
            250.6                  \\
                                   &
            ---                    & 
            \LQCD                  &
            31.50                  &
            60                     &
            99.91                  &
            240.8                  \\
        \midrule            
            3/2/1                  & 
            ---                    & 
            \LQCD + \QCDSR         &
            66.87                  &
            130                    &
            99.99                 &
            355.6                  \\
        \bottomrule
    \end{tabular}
    \caption{%
        \label{tab:results:summary}
        Summary of the outcomes of the various fits. We show all five fit models (two different truncations $\otimes$ usage of \SUF symmetry)
        to the full lattice QCD data set.
        We note that since none of the unitarity bounds are saturated at the best-fit points, they are excluded from the effective d.o.f. count.
    }
\end{table}

We apply all five fit models to the full lattice QCD data set.
For phenomenological purposes, we draw posterior samples for all of these analyses.
The posterior samples constitute one of the main numerical results of this work.
We provide them as part of the supplementary material~\cite{EOS-DATA-2025-02}.
All five fits are of excellent quality, as can be seen from the goodness-of-fit criteria
listed in \cref{tab:results:summary}. All $p$ values are in excess of $99\%$, indicating 
difficulties for a straight-forward statistical interpretation. Possible explanations for these large $p$ values, discussed in more detail below, include overestimated theory uncertainties in some or all of the fitted data sets; a dilution of the $p$ value due to the large number of observations; or systematic effects inherent to the use of pseudodata points in all of the analysed data sets.
As a consequence, we do not consider the overall $p$ value to be a useful criterion of fit quality and investigate the
individual analyses at a more ``microscopic'' level.
In particular, we investigate individual contributions to the overall $\chi^2$, which we use to abbreviate
the following expressions:
\begin{equation}
    \chi^2 = -2 \ln P(D \,|\, \vec{x} = \vec{x}^*, M)\,,
\end{equation}
where $\vec{x}^*$ denotes the mode of the posterior $P(\vec{x} \, |\, D, M)$.
Note that at $\vec{x}^*$ all unitarity bounds are respected, so the penalty function evaluates to unity.
\\

We begin this investigation at hand of the $2/1/0^*$ model with usage of \SUF symmetry, for two reasons.
First, its analysis leads to the largest $\chi^2 = 52.79$ value seen for the \LQCD data set.
Second, it is the most predictive of our fit models, since it features the smallest number of parameters,
thereby explaining why it leads to the largest $\chi^2$ value among all of our analyses of the \LQCD data set.\\
The largest contribution to its $\chi^2$ value arises from the \BToDstar \HPQCD[~2023] data set, which contributes $18.96$ with $56$ observations.
Based on these quantities, we can compute a ``local $p$ value'' that does not account for a reduction in the degrees of freedom
due to the number of free fit parameters. The local $p$ value for this particular data set is also in excess of $99\%$, reflecting
the overall quality of the fit.
We continue with finding the smallest individual local $p$ value amongst all constraints entering the data set. We find the two smallest local $p$ values
to be $16.4\%$ and $39.7\%$, which arise in the ratio of \BsToDs and \BToD tensor form factors over their respective vector form factors, respectively.
These results further solidify the picture of an excellent description of the data.\\
Nevertheless, we are concerned about the possibility that the local $p$ values for the three individual \BToDstar data sets might be
diluted due to the large number of observations. To investigate if such a dilution is present, we proceed with
a diagonalization of their respective $\chi^2$ contributions through an eigenvalue decomposition; see \cref{app:chisquare-diagonalisation}
for a description. The procedure produces individual local $p$ values for each independent linear combination
of the form factor data points. Among all of these local $p$ values, we find the smallest $p$ value to be $11.9\%$;
it arises in the description of the \HPQCD[~2023] data set and corresponds to a linear combination of the $\BToDstar$ and $\BsToDsstar$
form factors that is dominated by $h_{T_1}^{\BsToDsstar}(q^2 = 7.95\,\GeV^2)$.
We conclude that the most predictive model ($2/1/0^*$ with usage of \SUF) provides an excellent description of the available data,
even at the level of the individual independent observations.\\
We repeat this analysis for the four other fit models.
As expected, our findings for these fit models are qualitatively the same, with none of the local $p$-values providing any level of concern.

\begin{figure}[t!]
    \includegraphics[width=\textwidth]{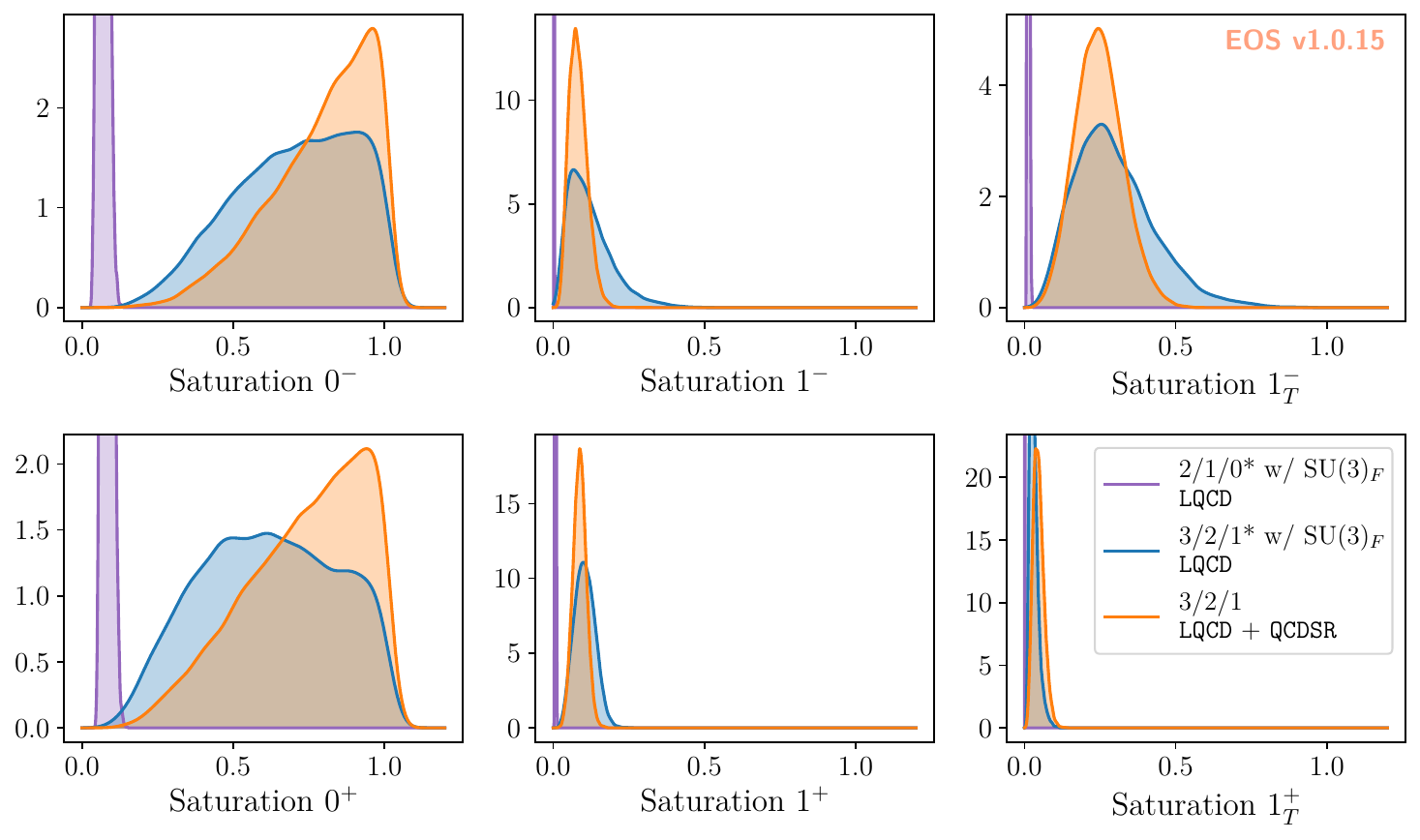}
    \caption{%
        \label{fig:results:summary:saturations} 
        Posterior predictive distributions for the unitarity bounds saturation for each $\bar{c}b$ current.
        The two plots on the right show the saturation of the tensor currents for the first time.
    }
\end{figure}

Based on these results, we find the fit model $2/1/0^*$ with \SUF to be adequate to describe the available data.
However, we follow the arguments discussed in Refs.~\cite{Bordone:2019guc,Bordone:2019vic}:
\begin{itemize}
    \item The power expansion of the form factors in $\varepsilon_Q$ does not translate into a power expansion of 
    the BGL coefficients at the same order. This problem is related to the decreasing size of the truncation order
    in $z$ as the power of $\varepsilon_Q$ increases.
    As a consequence, we use for the $2/1/0^*$ models $N_\text{UB} = 1$ while in the $3/2/1^{(*)}$ models we take one additional
    term in the bounds into account, $N_\text{UB} = 2$.
    Hence, the impact of the unitarity bounds is more accurately accounted for in the $3/2/1^{(*)}$ models. 
    \item Most importantly, using the $2/1/0^*$ fit model with or without \SUF has the potential to underestimate systematic theoretical uncertainties inherent in the \HQE. The proposed procedure in Refs.~\cite{Bordone:2019guc,Bordone:2019vic} is therefore to establish the most predictive fit model with an
    acceptable $p$ value and to then increase the truncation order in $z$ by one unit for all Isgur-Wise functions.
\end{itemize}
Consequently, we choose the fit model $3/2/1^*$ w/ \SUF to be our nominal fit model and use it
to derive our nominal numerical results presented in the remainder of this article.

One concern in the application of the \HQE is the convergence of the power expansion.
Although one cannot make all-order statements based on the first few coefficients in this expansion,
it is possible to exclude the \HQE as a useful tool if these first few coefficients are excessively large.
Of particular interest is the size of the NNLP parameters $\ell_{iq}(1)$ and $\ell_{iq}^{(1)}(1)$.
In the $2/1/0^*$ models, we find all NNLP parameters to be $\mathcal{O}(1)$ with matching uncertainties.
In the $3/2/1^*$ models, the picture is similar, with one exception: we find $\ell_{2d}^{(1)}(1) \simeq 20 \pm 20$.
While this is noteworthy, it does not indicate a failure of the \HQE as such. This slope is compatible
with zero, and a prefactor $(w - 1)$ in the expansion of $\ell_{2d}(w)$ further suppresses the impact of this
substantial parameter value in the semileptonic phase space.
We conclude that, in light of the available \LQCD data, the \HQE is free of pathological behaviour up to order $\varepsilon^2$.

Having investigated the convergence of the \HQE, we turn to the unitarity bounds.
We show the individual saturations of the six different unitarity bounds in
\cref{fig:results:summary:saturations}, depicting the outcomes of three different analyses:
(a) fit model $2/1/0^*$ w/ \SUF using the \LQCD dataset;
(b) fit model $3/2/1^*$ w/ \SUF using the \LQCD dataset;
and (c) fit model $3/2/1$ using the combined \LQCD + \QCDSR dataset.
This illustrates the effects the bounds have on the most predictive, the nominal, and the
least predictive of our fit models.
As expected based on previous analyses~\cite{Bordone:2019vic, Bordone:2019guc},
we find the largest effects due to the bounds in the scalar and pseudoscalar
channels for the $3/2/1$ model. The constraints in the vector and axial channels
have negligible impact, as does the one in the new $1^+_T$ channel.
In contrast, the saturation of the new $1^-_T$ channel reaches an average of $\simeq 25\%$
in the $3/2/1^{(\ast)}$ family of fit models.
While this is not a phenomenological relevant result on its own, it leaves open the question if
accounting for one-body hadronic contributions or
a joint analysis with form factors for other processes --- like $\bar{B}_c\to J/\psi$, $\Lambda_b \to \Lambda_c$ or $\bar B_{(s)}\to D_{(s)}^{**}$ --- might increase the saturation to a point where
the parameter space is affected with phenomenological implications.

\subsection{Phenomenological predictions}
\label{sec:results:posterior-predictions}

\begin{figure}[t!]
    \centering
    \subfloat
    {
        \includegraphics[width=0.45\textwidth]{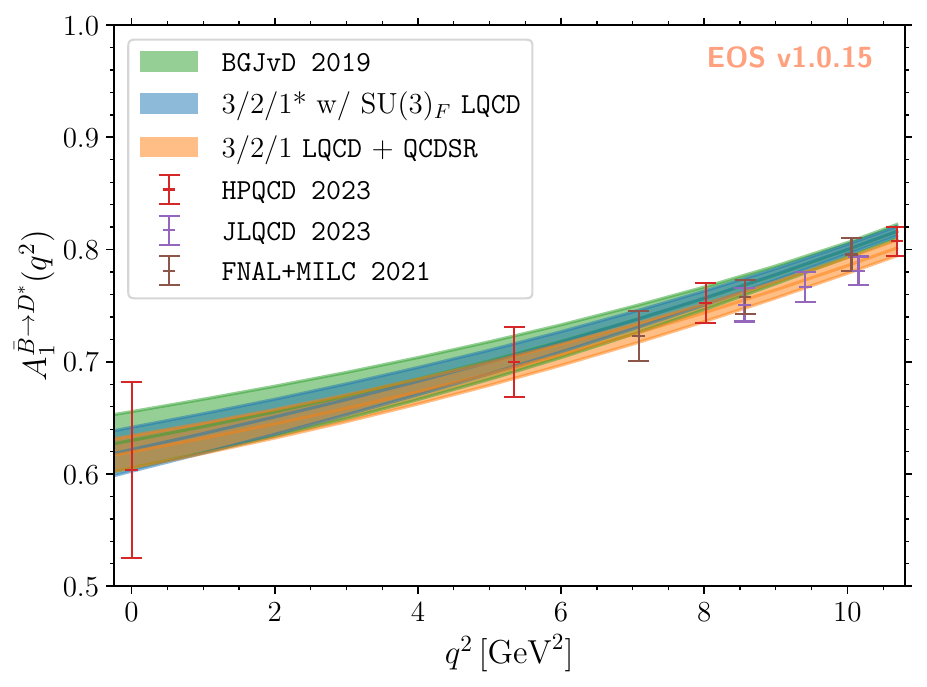}
    }
    \hfill
    \subfloat
    {
        \includegraphics[width=0.45\textwidth]{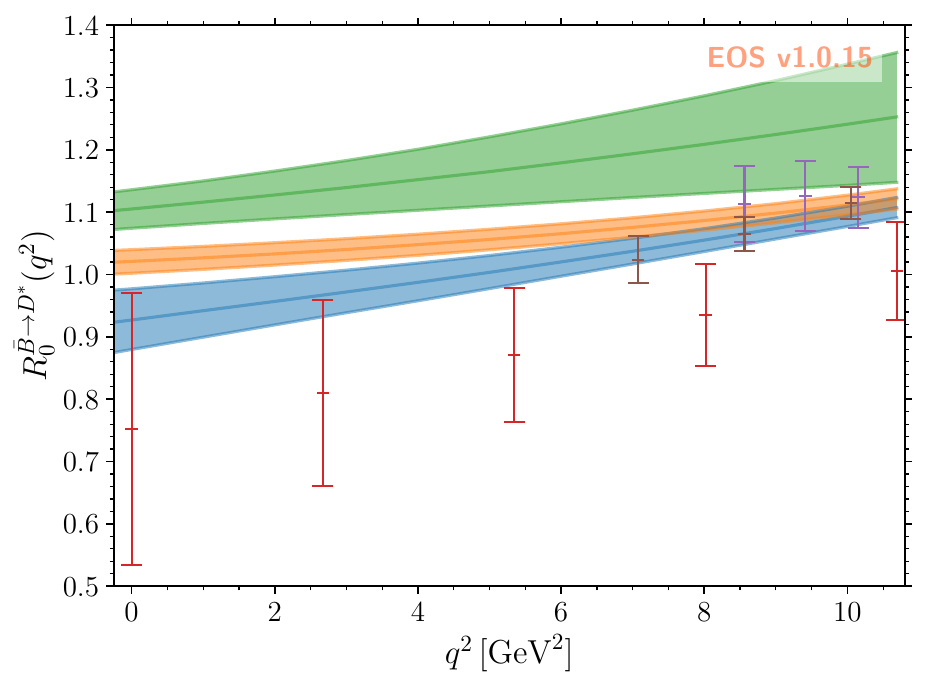}
    }
    
    
    \subfloat
    {
        \includegraphics[width=0.45\textwidth]{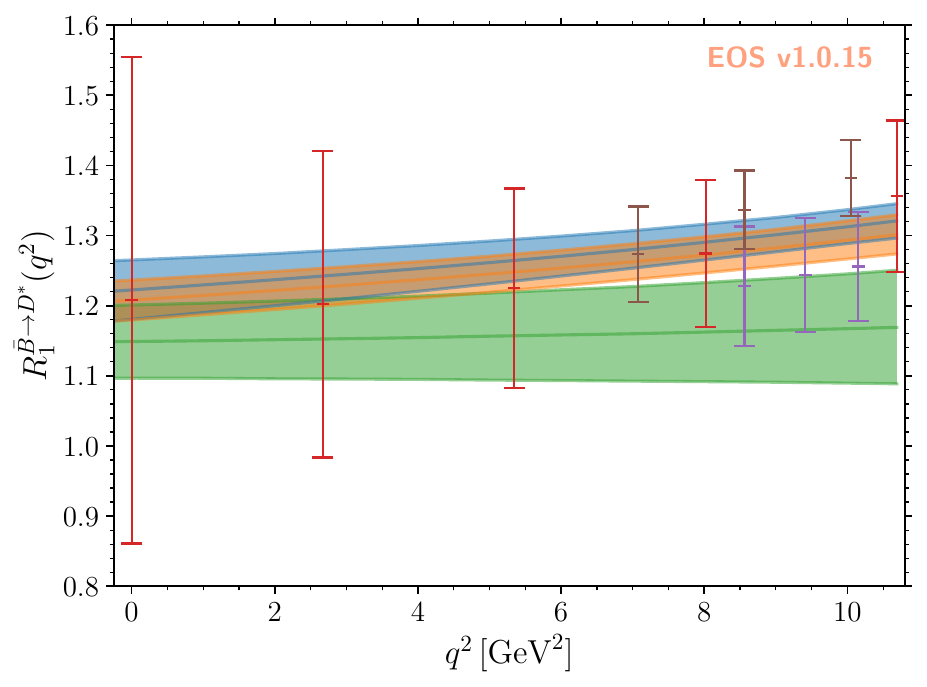}
    }
    \hfill
    \subfloat
    {
        \includegraphics[width=0.45\textwidth]{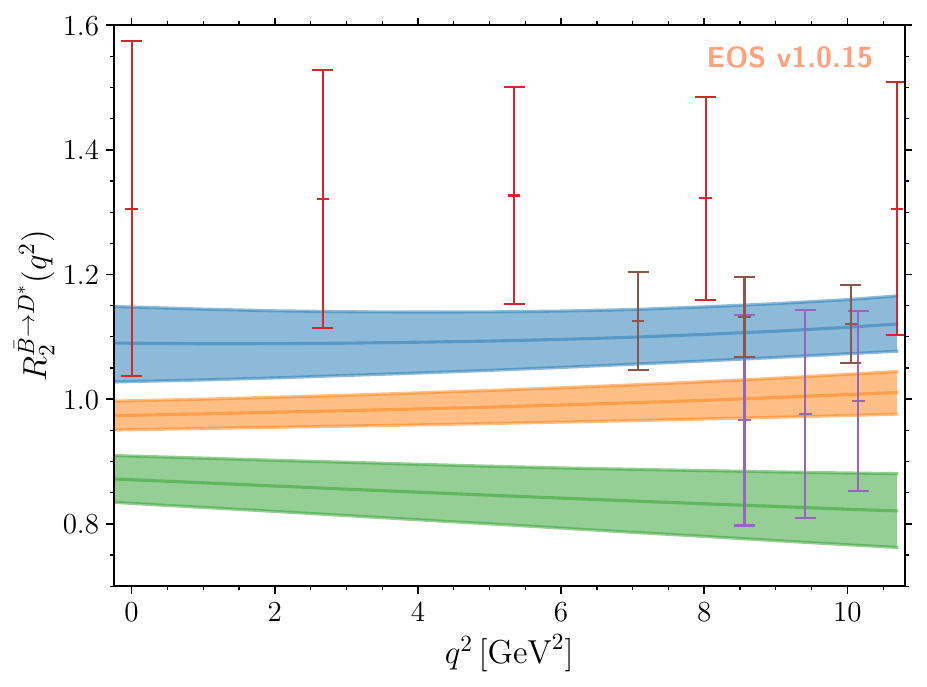}
    }
    
    \caption{%
        Plot of the \BToDstar form factor $A_1$ and the form factor ratios $R_{0,1,2}$, all as functions of $q^2$.
        Solid curves correspond to the median values and shaded regions correspond to the central $68\%$ credible interval.
        The nominal results of our previous work involving sum-rule results are shown in green.
        Our nominal results of this work in the $3/2/1^*$ model with $\SUF$ symmetry are shown in blue.
        Auxiliary results of this work in the $3/2/1$ model are shown in orange.
    }
    \label{fig:results:comparison-A1-and-ratios}
\end{figure}

The posterior samples for the \HQE parameters obtained in \cref{sec:results:summary}
do not follow a multivariate Gaussian distribution.
Hence, providing their sample mean and covariance matrix is not advisable, since their use
in numerical predictions would lead to biases for and misestimations of the derived quantities.
Here, we use these posterior samples to produce posterior-predictive distributions
for a variety of derived quantities with phenomenological relevance.
They are also part of the main numerical results of our work.

\paragraph{Form factor ratios}
As a first step, we predict the form factor $A_1$ and the form factor ratios $R_0$, $R_1$, and $R_2$ for the $\BToDstar$ transitions.
In our choice of form factor basis, these ratios are defined as~\cite{Faller:2008tr}
\begin{align}
    R_0(q^2) &= \left(1 - \frac{q^2}{t_+^{BD^*}}\right)\! \frac{A_0}{A_1}
    , &
    R_1(q^2) &= \left(1 - \frac{q^2}{t_+^{BD^*}}\right)\! \frac{V}{A_1}
    , &
    R_2(q^2) &= \left(1 - \frac{q^2}{t_+^{BD^*}}\right)\! \frac{A_2}{A_1}
    .
\end{align}
In \cref{fig:results:comparison-A1-and-ratios}, we compare our nominal posterior predictions in the $3/2/1^*$ model with the application of $\SUF$ symmetry for the \LQCD-only likelihood
with our posterior predictions in the $3/2/1$ model for the combined \LQCD + \QCDSR likelihood. We further compare to the nominal posterior predictions
of our previous work~\cite{Bordone:2019vic}.
The plot of the form factor $A_1$ reflects the excellent agreement of the individual \LQCD and \QCDSR likelihoods within the fit, showing only small variations across the different fits and
overlap of the uncertainty envelopes.
The plot for the form factor ratio $R_1$ indicates systematically different trends between the outcome
of our previous analysis (green) and the outcomes in this work (blue and orange).
Nevertheless, there is no significant tension between the two trends, dominantly due to the large
uncertainties emerging from the previous analysis.
In contrast, the visual comparison of the two remaining form factor ratios $R_0$ and $R_2$ shows striking differences.
This seems to suggest significant tensions, which are not corroborated at the level of the parameters' test statistics.
Our findings appear to be compatible with those of Ref.~\cite{Martinelli:2023fwm,Bordone:2024weh}
in the context of fits within the BGL parametrization and without application of the \HQE,
further fuelling the ongoing discussion in the community.
The very large $p$ values observed here appear to us to be indicative of issues in the use of pseudodata;
we intend to investigate this issue elsewhere.
\\

\paragraph{Lepton-flavour universality ratios}
\label{sec:results:posterior-predictions:lfu}
As a second step, we predict the lepton flavour universality (LFU) ratios $R(D_{(s)}^{(*)})$
\begin{align}
    R(D_{(s)}^{(*)}) &= 
    \frac{
        \mathcal{B}(\bar B_{(s)} \to D_{(s)}^{(*)} \tau^- \bar\nu_\tau)
    }{
        \mathcal{B}(\bar B_{(s)} \to D_{(s)}^{(*)} \ell^- \bar\nu_\ell)
    }\,,
    \qquad
    \text{with }\ell=e,\mu\,.
\end{align}
In our nominal fit model $3/2/1^*$ w/ \SUF, using the \LQCD likelihoods, we find
\begin{equation}
\begin{aligned}
    R(D)          
        & = 0.3022 \pm 0.0036\,,   &
    R(D^*)        
        & = 0.2589 \pm 0.0042\,,   \\
    R(D_s)        
        & = 0.2990 \pm 0.0031\,,   &
    R(D_s^*)      
        & = 0.2605 \pm 0.0047\,.
\end{aligned}
\end{equation}
The results above are approximately distributed as a multivariate Gaussian with the correlation matrix
\begin{equation}
\newcommand{\pp}{\phantom{+}}
\begin{pmatrix}
    \pp 1.0000 & \pp 0.0116 & \pp 0.1575 & \pp 0.0146 \\
    \pp 0.0116 & \pp 1.0000 & \pp 0.0119 & \pp 0.3361 \\
    \pp 0.1575 & \pp 0.0119 & \pp 1.0000 & \pp 0.0261 \\
    \pp 0.0146 & \pp 0.3361 & \pp 0.0261 & \pp 1.0000         
\end{pmatrix}\,,
\end{equation}
with the ordering $R(D)$, $R(D^*)$, $R(D_s)$, and $R(D_s^*)$.
Our result for $R(D^*)$ is in excellent agreement with the results of Refs.~\cite{Martinelli:2023fwm,Bordone:2024weh} using the same lattice data sets.
However, our results for $R(D_{(s)}^*)$ are about $2\sigma$ above our previous analysis \cite{Bordone:2019guc} as well as analyses using experimental $\BToDstar$ data, see Ref.~\cite{HeavyFlavorAveragingGroupHFLAV:2024ctg} for on overview.\\

In the fit model $3/2/1$, using the \LQCD + \QCDSR likelihoods, we find
\begin{equation}
\begin{aligned}
    R(D)          
        & = 0.2983 \pm 0.0031\,,   &
    R(D^*)        
        & = 0.2510 \pm 0.0026\,,   \\
    R(D_s)        
        & = 0.2989 \pm 0.0041\,,   &
    R(D_s^*)      
        & = 0.2508 \pm 0.0029\,.
\end{aligned}
\end{equation}
Their correlation matrix reads
\begin{equation}
\newcommand{\pp}{\phantom{+}}
\begin{pmatrix}
    \pp 1.0000 & \pp 0.0855 & \pp 0.0712 & \pp 0.0095 \\
    \pp 0.0855 & \pp 1.0000 & -   0.0100 & \pp 0.2903 \\
    \pp 0.0712 & -   0.0100 & \pp 1.0000 & \pp 0.0006 \\
    \pp 0.0095 & \pp 0.2903 & \pp 0.0006 & \pp 1.0000         
\end{pmatrix}\,.
\end{equation}
Our result for $R(D^*)$ shows a sizeable deviation with respect to the results of Refs.~\cite{Martinelli:2023fwm,Bordone:2024weh} and our nominal fit, while agreeing with our previous analysis and values obtained including experimental data.
It is not useful to assign a statistical significance to the tension with Refs.~\cite{Martinelli:2023fwm,Bordone:2024weh}, since our prediction and the references' predictions
are dominantly relying on the same underlying lattice QCD results.

\paragraph{Angular observables}
The decays $\BToDstar\ell^-\bar\nu$ have phenomenologically interesting angular distributions~\cite{Blanke:2018yud,Alguero:2020ukk,Bobeth:2021lya,Fedele:2022iib,Fedele:2023ewe,Martinelli:2024vde,Martinelli:2024bov}.
For what follows, we use the conventions of Ref.~\cite{Bobeth:2021lya}.
Making predictions for all angular distributions and observables is impractical.
Here, we present results only for a limited set of particularly interesting angular observables arising in these distributions.
Further predictions can be readily obtained from the \EOS software~\cite{EOS:v1.0.15}, using the
provided posterior samples~\cite{EOS-DATA-2025-02} as input.

The distribution in the helicity angle of the $D\pi$ system is dominated by the so-called longitudinal polarisation
$F_L$. For our nominal analysis (fit model $3/2/1^*$ w/ \SUF, \LQCD likelihoods), we obtain
\begin{equation}
\begin{aligned}
    F_L^{\BToDstar e^-\bar\nu}     & = 0.483 \pm 0.012  &
    F_L^{\BsToDsstar e^-\bar\nu}   & = 0.473 \pm 0.014  \\
    F_L^{\BToDstar\mu^-\bar\nu}    & = 0.483 \pm 0.012  &
    F_L^{\BsToDsstar\mu^-\bar\nu}  & = 0.473 \pm 0.014  \\
    F_L^{\BToDstar\tau^-\bar\nu}   & = 0.4300 \pm 0.0062    &
    F_L^{\BsToDsstar\tau^-\bar\nu} & = 0.4285 \pm 0.0072    \,.
\end{aligned}
\end{equation}
Our result for $F_L^{\BToDstar\tau^-\bar\nu}$ is in excellent agreement with the results of Refs.~\cite{Martinelli:2023fwm,Bordone:2024weh}, as well as with our previous analysis \cite{Bordone:2019guc}.\\

For fit model $3/2/1$ and the combined \LQCD + \QCDSR likelihood, we obtain 
\begin{equation}
\begin{aligned}
    F_L^{\BToDstar e^-\bar\nu}     & = 0.5099 \pm 0.0062    &
    F_L^{\BsToDsstar e^-\bar\nu}   & = 0.5090 \pm 0.0092    \\
    F_L^{\BToDstar\mu^-\bar\nu}    & = 0.5101 \pm 0.0062    &
    F_L^{\BsToDsstar\mu^-\bar\nu}  & = 0.5092 \pm 0.0092    \\
    F_L^{\BToDstar\tau^-\bar\nu}   & = 0.4448 \pm 0.0040    &
    F_L^{\BsToDsstar\tau^-\bar\nu} & = 0.4453 \pm 0.0062    \,.
\end{aligned}
\end{equation}
As in the case of $R(D^*)$, our posterior prediction for $F_L^{\BToDstar\tau^-\bar\nu}$ shows a sizeable deviation with respect to the results of Refs.~\cite{Martinelli:2023fwm,Bordone:2024weh} and our nominal fit, but also with our previous analysis; we refer back to our discussion in the LFU paragraph.

\begin{figure}[t!]
    \centering
    \includegraphics[width=0.666\textwidth]{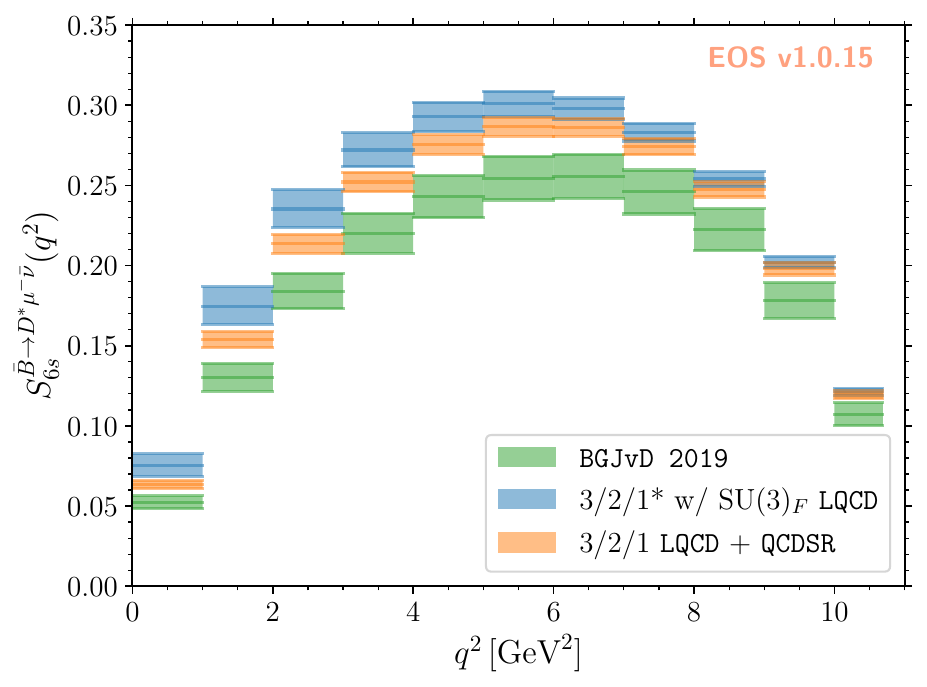}
    \caption{%
        \label{fig:results:summary:S6s} 
        Plot of the angular observable $S_{6s}$ for the $\bar B \to D^* \mu^- \bar\nu_\ell$ decays in bins of $q^2$.
    }
\end{figure}
The distribution in the helicity angle of the $\ell^-\bar\nu$ system is dominated by the so-called forward-backward
asymmetry $A_{FB}$. In the SM, this observable can be identified with the angular observable $S_{6s}$.
As an example, we present in \cref{fig:results:summary:S6s} a plot of $S_{6s}$ in $\BToDstar\mu^-\bar\nu$,
integrated over five bins in $q^2$, for the various posterior choices.

\paragraph{BGL coefficients}
Although the posterior distributions of the \HQE parameters are not Gaussian,
we find that the posterior-predictive distributions for the BGL coefficients derived from the \HQE posterior
can be well approximated by a multivariate Gaussian distribution.
We provide the sample mean and the sample covariance matrix for the \BToD, \BToDstar, \BsToDs, and \BsToDsstar BGL coefficients separately as part of the supplementary material~\cite{EOS-DATA-2025-02}. These results are produced from our nominal fit in the $3/2/1^*$ model with \SUF.
We remove the zeroth-order coefficients of the form factors $f_0$, $A_0$, $A_{12}$, $T_2$, and $T_{23}$ from our results,
since these coefficients are not independent from the remaining BGL coefficients on account of
the endpoint relations between these form factors.
This leaves us with $8$ independent \BqToDq and $17$ independent \BqToDqStar BGL coefficients per spectator quark flavour $q$.
Given that in our nominal fit we only have $16$ independent parameters entering each \BqToDqStar transition,
we have to remove one \BqToDqStar BGL coefficient from our posterior predictions to ensure that the covariance matrix remains regular.
We remove the order $z^2$ coefficient of the $T_{23}$ form factor in the \BToDstar and \BsToDsstar processes,
which is neither needed in the SM predictions nor does it contribute dominantly to the tensor amplitudes in BSM scenarios.

\section{Conclusion}

We have performed a comprehensive analysis within the \HQE of the available lattice QCD results for \BqToDqDqStar form factors,
as published by the FNAL/MILC, HPQCD, and JLQCD collaborations~\cite{Bailey:2012rr,MILC:2015uhg,Na:2015kha,McLean:2019qcx,FermilabLattice:2021cdg,Harrison:2023dzh,Aoki:2023qpa}.
Our analysis closely follows our previous works~\cite{Bordone:2019vic,Bordone:2019guc} that have crucially relied on QCD sum rule inputs.
Given the variety of the now-available lattice QCD inputs, reliance on these QCD sum rules inputs is no longer necessary
to achieve an \HQE description of the form factors.
Employing the full phenomenological power of the heavy-quark spin symmetry, we derive for the first time strong unitarity
bounds for the two tensor currents. \\

To determine if the recent lattice QCD results are well described within the \HQE, we have performed a joint fit in five distinct fit models.
These models feature two different truncation orders for the parametrization of the Isgur-Wise functions and the possibility
to share the next-to-next-to-leading power (NNLP) Isgur-Wise function.
For all five models, we obtain $p$ values in excess of $99\%$.
This result assumes that the systematic lattice QCD uncertainties can, at least to some extent, be interpreted in a statistical fashion.
Moreover, we find that fit models describe well each individual results by the FNAL/MILC, the HPQCD, and the JLQCD collaborations.
We therefore conclude that the individual lattice QCD results are mutually compatible.
We further find that the convergence of the \HQE is not hampered by overly large expansion coefficients.
In particular, we find all \HQE parameters to be compatible with the naive expectation of being $\mathcal{O}(1)$ parameters.
The interpretation of these results requires some care, however, given the use of an HQE-like parametrization within the available lattice QCD analyses.
Including the tensor form factors in the fit provides important complementarity constraints on the \HQE parameters.
We therefore encourage the inclusion of tensor form factors in future lattice QCD analyses of \BToDDstar form factors.
Given the strong correlations between \BToD and \BToDstar form factors within the \HQE parametrizations due to heavy-quark spin symmetry,
a joint lattice QCD analysis of both modes will further help constraining the \HQE parameters.
\\

We predict a variety of physical observables of $\BqToDqDqStar\tau^-\bar\nu$ decays, including angular observables and LFU ratios.
Our nominal results
feature somewhat smaller uncertainties than obtained from a BGL analysis of the same data~\cite{Bordone:2024weh},
and significantly smaller uncertainties than obtained from a dispersive matrix analysis of the same data~\cite{Martinelli:2023fwm}.
This is not surprising, since the \HQE fit model is more predictive than a BGL fit or a dispersive matrix analysis.\\
When comparing our posterior predictions with those based on our 2019 results~\cite{Bordone:2019guc,Bordone:2019vic},
we surprisingly find large gaps that visually suggest an incompatibility.
This observation is most striking in the form factor ratios $R_0$ and $R_2$.
To determine if these gaps are statistically significant, we have performed a joint fit to the
QCD sum rule inputs used in the 2019 analysis and the full set of lattice QCD results available now.
The joint fit features a $p$ value in excess of $99\%$, indicating full compatibility amongst the inputs when taken at face-value.\\
We note in passing that the numerators of the ratios $R_0$ and $R_2$ are comprised of the form factors $A_0$ and $A_{12}$,
to which lattice QCD analyses lose sensitivity as they approach the zero-recoil point $q^2 \to q^2_\text{max}$.
Unfortunately, these form factors are also very relevant to phenomenological applications.
We look forward to the results of ongoing and future lattice QCD analyses, which have the potential
to probe these two form factors with more precision and therefore to clarify the current situation.
The strong variability in the predictions, despite the overall good agreement as indicated by the $p$ value,
poses a problem for the phenomenological application of the lattice data.
We wonder if, following precedent set in the investigation of theory predictions for the hadronic contributions to $(g-2)_\mu$,
it could be useful to define a suitable set of diagnostic quantities that can be more directly compared by the lattice QCD collaborations,
as previously suggested in Ref.~\cite{Tsang:2023nay}.
This could help uncover potential sources of problems.
\\

Within the framework of the \HQE, the combination of the available lattice QCD results
respects the unitarity bounds, including our newly derived strong unitarity bounds for the two tensor currents.
This corroborates the findings of Ref.~\cite{Bordone:2024weh} using a more predictive fit model.
While we find the new bound for the $1^+$ tensor current to be of no phenomenological
relevance, we find a substantial $\mathcal{O}(25\%)$ saturation of the bound for the $1^-$ tensor current; significantly
larger than the saturations of the bounds for the $1^-$ and $1^+$ (axial)vector currents.
We conjecture that a simultaneous analysis with further exclusive $b\to c$ tensor form factors
has the potential to put phenomenological bounds on the \HQE parameter space.

\section*{Acknowledgements}

\sloppy
NG thanks Marco Fedele for comments on the manuscript.
MB was partially supported by the Deutsche Forschungsgemeinschaft (DFG, German Research Foundation) under grant 396021762 - TRR 257. NG has been been partially supported by the UK Science and Technology Facilities Council (STFC)
through the consolidated grants ST/T000694/1 and ST/X000664/1.
The work of MJ is supported in part by the Italian Ministry of University and Research (MUR) under grant PRIN 2022N4W8WR.
DvD acknowledges support by the STFC
through the grants ST/V003941/1 and ST/X003167/1.

\appendix

\section{Relations between the form factor bases}
\label{app:form-factor-decomposition}

The change of basis that connects the QCD (\cref{eq:BtoD-trad1,eq:BtoDst-trad1,eq:BtoDst-trad2,eq:BtoDst-trad3,eq:BsttoD-trad1,eq:BsttoD-trad2,eq:BsttoD-trad3,eq:BstToDst-trad1,eq:BstToDst-trad2}) and HQE (\cref{eq:BtoDtensor1,eq:BtoDstensor1,eq:BtoDstensor2,eq:BstoDtensor1,eq:BstoDtensor2,eq:BstoDstensor1,eq:BstoDstensor2}) bases reads:
\begin{align*} 
    \frac{1}{\sqrt{\mB\mD}}\, 
    h_{T} 
    &= \frac{2}{m_B+m_D}  f_T\,,\\
    \frac{1}{\sqrt{\mB\mDs}}\, 
    h_{T_1} 
    &= - \frac{\mBs+m_D}{2 q^2 \mB \mDs}\left(s_-\,T_1 - (\mB-\mDs)^2\,T_2\right) \,,\\
    \frac{1}{\sqrt{\mB\mDs}}\, 
    h_{T_2} 
    &= \frac{\mDs-\mB}{2 q^2\mBs\mD}\left(s_+ T_1-(\mB+\mDs)^2 T_2\right)  \,,\\
    \frac{1}{\sqrt{\mB\mDs}}\, 
    h_{T_3} 
    &= -\frac{2 \mB}{q^2}\left(T_1-T_2 - \frac{q^2}{\mB^2-\mDs^2} T_3\right) \,,\\
    \frac{1}{\sqrt{\mBs\mD}}\, 
    h_{\bar{T}_1} 
    &=
    - \frac{\mBs+m_D}{2 q^2 \mBs m_D}\left(s_-\,\bar{T}_1 - (\mBs-\mD)^2\,\bar{T}_2\right) \label{eq:FF_basis_1}
    \,,\\
    \frac{1}{\sqrt{\mBs\mD}\, }
    h_{\bar{T}_2} 
    &=
    \frac{\mBs-\mD}{2 q^2\mBs\mD}\left(s_+ \bar{T}_1-(\mBs+\mD)^2 \bar{T}_2\right) 
    \,, \\
    \frac{1}{\sqrt{\mBs\mD}}\, 
    h_{\bar{T}_3} 
    &=
     -\frac{2 \mD}{q^2}\left(\bar{T}_1-\bar{T}_2 - \frac{q^2}{\mD^2-\mBs^2} \bar{T}_3\right)
    \,,\\
    \frac{1}{\sqrt{\mBs\mDs}}\, 
    h_{T_4}
    &=
    \frac{1}{\sqrt{\mBs\mDs}}T_4
    \,,\\
    \frac{1}{\sqrt{\mBs\mDs}}\, 
    h_{T_5}
    &=
    \frac{s_+(\mBs^2+\mDs^2-q^2)}{\sqrt{\mBs\mDs}\lBsDs} T_4
    -
     \frac{\mBs\mDs^2(\mBs^2-\mDs^2+q^2)}{q^2\lBsDs} T_5
    +
     \frac{\mDs}{q^2}T_9
    \,,
    \\
    \frac{1}{\sqrt{\mBs\mDs}}\, 
    h_{T_6}
    &=
    \frac{s_+(\mBs^2+\mDs^2-q^2)}{\sqrt{\mBs\mDs}\lBsDs}T_4
    -
    \frac{\mBs^2\mDs(\mBs^2-\mDs^2-q^2)}{q^2\lBsDs}T_6
    +
    \frac{\mBs}{q^2}T_8
    \,, \\
    \frac{1}{\sqrt{\mBs\mDs}}\, 
    h_{T_7}
    &=
    \frac{1}{\sqrt{\mBs\mDs}}T_7
    \,, \\
    \frac{1}{\sqrt{\mBs\mDs}}\, 
    h_{T_8}
    &=
    2\frac{s_+\sqrt{\mBs\mDs} }{\lBsDs}T_4
    -
    \frac{
    \mBs\mDs^2(\mBs^2-\mDs^2+q^2)}{q^2\lBsDs}T_6
    +
    \frac{\mDs}{q^2}T_8
    \,, \\
    \frac{1}{\sqrt{\mBs\mDs}}\, 
    h_{T_9}
    &=
    -2\frac{s_+\sqrt{\mBs\mDs}}{\lBsDs}T_4
    +
    \frac{\mBs^2\mDs(\mBs^2-\mDs^2-q^2)}{q^2\lBsDs}T_5
    -
    \frac{\mBs}{q^2} T_9
    \,, \\
    \frac{1}{\sqrt{\mBs\mDs}}\, 
    h_{T_{10}}
    &=
    \frac{2}{q^2\lBsDs}
    \bigg(q^2s_+\sqrt{\mBs\mDs} T_4+\mBs^2\mDs^2(\mDs T_5-\mBs T_6)\nonumber\\
    & -\sqrt{\mBs\mDs} q^2(\mBs^2+\mDs^2-q^2) T_7+\mBs^2\mDs(\mBs^2-\mDs^2-q^2) T_8\nonumber\\
    &
    -\mBs\mDs^2(\mBs^2-\mDs^2+q^2) T_9+2\mBs^{3/2}\mDs^{3/2}q^2T_{10}
    \bigg)
    \,. 
\end{align*}

\section{Diagonalisation of the multivariate constraints}
\label{app:chisquare-diagonalisation}

To determine if there is any accidental dilution of the local $p$ values discussed in \cref{sec:results:summary},
we investigate the multivariate constraints using the following diagonalization procedure.
Let $\vec{p}$ be the vector of form-factor predictions at the best-fit point, and
let $\vec{d}$ be the vector of form-factor determinations by the lattice QCD and QCD sum rule analyses.
Let $\Sigma$ be the covariance matrix of these determinations.
Since $\Sigma$ is a regular and positive definite matrix, we can diagonalise the $\chi^2$ contribution
as
\begin{equation}
    \chi^2 = (\vec{p} - \vec{d})^T\, \Sigma^{-1}\, (\vec{p} - \vec{d})
           = \sum_{n=1}^{N = \dim \vec{d}} \lambda_n^{-1} \left|\left[D (\vec{p} - \vec{d})\right]_n\right|^2
           \equiv \sum_n^N \chi_n^2\,.
\end{equation}
Here, $\Sigma = D \Lambda D^{-1}$ with $\Lambda = \diag(\lambda_1, \dots, \lambda_N)$
and with orthogonal matrix $D$.\\

In extreme cases, a single contribution (\eg, $\chi_1^2$) could dominate the entire $\chi^2$ value,
(\ie, $\chi_1^2 \simeq \chi^2$).
Such a case can arise, \eg, if a single determination is less correlated
to the rest of the determinations than the rest of the determinations to each other.
In such a case, the $p$ value for the total $\chi^2$ with $N$ degrees of freedom might well be acceptable (\ie, above our a-priori threshold of $3\%$),
while the $p$ value for the largest single contribution $\chi^2_1$ with one degree of freedom might not be.
As discussed in \cref{sec:results:summary}, we find that no such case arises in our fits.

\bibliographystyle{JHEP}
\bibliography{references}

\end{document}